\documentclass[a4paper,11pt]{article} 
\usepackage[a4paper]{geometry}
\usepackage{authblk}
\usepackage{graphicx}
\usepackage{amssymb}
\usepackage{amsmath}
\usepackage{url}
\usepackage{bm}
\usepackage{subfig}
\usepackage{tabularx}
\usepackage{tikz}
\usepackage{pgfplots}
\usepackage{float}
\usepackage{multirow}
\usepackage{verbatim}
\usetikzlibrary{arrows,intersections}
\usepackage[europeanresistors,americaninductors]{circuitikz}

\usepackage{subfig}
\usepackage{stackengine}

\title{Active network management for electrical distribution systems: problem formulation, benchmark, and approximate solution}

\author[*]{Quentin Gemine}
\author[*]{Damien Ernst}
\author[*]{Bertrand Corn{\'e}lusse}

\affil[*]{\{\texttt{qgemine@ulg.ac.be}, \texttt{dernst@ulg.ac.be}, \texttt{bertrand.cornelusse@ulg.ac.be}\}\\Department of Electrical Engineering and Computer Science, University of Li\`ege, 4000 Li\`ege, Belgium}

\date{May 31, 2016}

\providecommand{\keywords}[1]{\textbf{\textit{Index terms---}} #1}

\begin{document}

\maketitle

\begin{abstract}
With the increasing share of renewable and distributed generation in electrical distribution systems,  Active Network Management (ANM) becomes a valuable option for a distribution system operator to operate his system in a secure and cost-effective way without relying solely on network reinforcement. ANM strategies are short-term policies that control the power injected by generators and/or taken off by loads in order to avoid congestion or voltage issues. While simple ANM strategies consist in curtailing temporary excess generation, more advanced strategies rather attempt to move the consumption of loads to anticipated periods of high renewable generation. However, such advanced strategies imply that the system operator has to solve large-scale optimal sequential decision-making problems under uncertainty. The problems are sequential for several reasons. For example, decisions taken at a given moment constrain the future decisions that can be taken,  and decisions should be communicated to the actors of the system sufficiently in advance to grant them enough time for implementation. Uncertainty must be explicitly accounted for because neither demand nor generation can be accurately forecasted. 
We first formulate the ANM problem, which in addition to be sequential and uncertain, has a nonlinear nature stemming from the power flow equations and a discrete nature arising from the activation of power modulation signals. This ANM problem is then cast as a stochastic mixed-integer nonlinear program, as well as second-order cone and linear counterparts, for which we provide quantitative results using state of the art solvers and perform a sensitivity analysis over the size of the system, the amount of available flexibility, and the number of scenarios considered in the deterministic equivalent of the stochastic program. To foster further research on this problem, we make available at \url{http://www.montefiore.ulg.ac.be/~anm/} three test beds based on distribution networks of 5, 33, and 77 buses. These test beds contain a simulator of the distribution system, with stochastic models for the generation and consumption devices, and callbacks to implement and test various ANM strategies. 
\end{abstract}

\maketitle

\section{Notation}
We present here the main elements of notation used throughout the text. Some locally defined notation may not be covered in this section.
\paragraph{Indices:} {~} \\
\begin{tabularx}{\textwidth}{r p{10cm}}
	$d$ & Device connected to a node.\\
	$m$ or $n$ & Node of the electrical system.\\
	$mn$ & Link of the electrical system between nodes $m$ and $n$.\\
	$t$ & Time period.\\
\end{tabularx}

\paragraph{Sets:} {~} \\
\begin{tabularx}{\textwidth}{r p{10cm}}
	$\mathcal{D}$ & Set of electrical devices.\\
	$\mathcal{G}$ & Subset of $\mathcal{D}$ containing distributed generators.\\
	$\mathcal{C}$ & Subset of $\mathcal{D}$ that are electrical loads.\\
	$\mathcal{F}$ & Subset of $\mathcal{C}$ that can be controlled by the DSO.\\
	$\mathcal{T}$ & Set of time periods.\\
	$\mathcal{N}$ & Set of nodes of the electrical system.\\
	$\mathcal{L}$ & Set of links of the electrical system.\\
	$\mathcal{S}^{(i)}_t$ & Space of state vector $\bm{s}^{(i)}_t$ (see the variables below for $i \in \{1,2,3\}$). \\ 
	$\mathcal{S}$ & Global state space of the system. \\ 
	$\mathcal{A}_{\bm{s}}$ & Feasible action (or control) space in state $s \in \mathcal{S}$. \\ 
	$\mathcal{A}_{d,\bm{s}}$ & Feasible set of $act_{d,t}$ (see variables).\\
	$\mathcal{W}$ & Set of possible realizations of random processes.\\
\end{tabularx}

\paragraph{Parameters:} {~} \\[1mm]
\begin{tabularx}{\textwidth}{r p{10cm}}
	$Y^{(br)}_{mn}$ & Branch admittance of link $(m,n)$.\\
	$Y^{(sh)}_{mn}$ & Shunt admittance of link $(m,n)$ on the side of node $m$.\\
	$t_{mn}$ & Transformation ratio of link $(m,n)$ on the side of node $m$. \\
	$\bm{Y}$ & Nodal admittance matrix.\\
	$\bm{Y}_{n\cdot}$ & $n^{\text{th}}$ row of $\bm{Y}$.\\
	$Y_{mn}$ & Element $(m,n)$ of $\bm{Y}$. \\
	$\underline{V_n}$ and $\overline{V_n}$ & Lower and upper operational limits on voltage magnitude $|V_n|$.\\
	$\overline{I}_{mn}$ & Operational limit on current magnitude $|I_{mn}|$.\\
	$\tan \phi_d$ & Reactive to active power ratio of device $d$ (assumed constant for all devices).\\
	$T_d$ & Duration of a modulation signal sent to a flexible load.\\
\end{tabularx}
\noindent\begin{tabularx}{\textwidth}{r p{10cm}}
	$\Delta P_d$ & Vector of length $T_d$ representing the modulation signal sent to a flexible load.\\
	$N_{loads}$ & Length of the history of load consumption tracked in the state. \\ 
	$N_{ir}$ & Length of the history of solar irradiance tracked in the state.\\ 
	$N_{v}$ & Length of the history of wind speed tracked in the state.\\ 
	$q_t$ & Index of a quarter of an hour.
\end{tabularx}

\paragraph{Variables:} Note that some variables may have an additional subscript $t$. Also some variables are control variables, some represent the state of the system, and the remaining variables are exogenous stochastic processes.\\
\begin{tabularx}{\textwidth}{r p{10cm}}
	$\bm{V}$ & Vector of size $|\mathcal{N}|$, node voltages.\\
	$V_n$ & Complex voltage at node $n$, i.e. $n^{\text{th}}$ component of $\bm{V}$.\\
	$\bm{I}$ & Vector of size $|\mathcal{N}|$, current injected in the nodes.\\
	$I_i$ & If $i=l$, it is the complex current in link $l$, if $i=n$, it is the complex current injected in bus $n$,  i.e. $n^{\text{th}}$ component of $\bm{I}$.\\
	$S_i$ & Apparent power injected in bus. If $i=d$, it is the power injected by device $d$. If $i=n$, it is the total power injected by all devices connected at node $n$.\\
	$P_i$ & Active power injected in bus. If $i=d$, it is the power injected by device $d$. If $i=n$, it is the total power injected by all devices connected at node $n$.\\
	$Q_i$ & Reactive power injected in bus. If $i=d$, it is the power injected by one device. If $i=n$, it is the total power injected by all devices connected at node $n$.\\
	$S_{mn}$ & Apparent power entering branch $l=(m,n)$ from the $m$ side.\\
	$P_{mn}$ & Active power entering branch $l=(m,n)$ from the $m$ side.\\
	$Q_{mn}$ & Reactive power entering branch $l=(m,n)$ from the $m$ side.\\
	$ir_t$ & Solar irradiance level at time $t$.\\
	$v_t$ & Wind speed at time $t$.\\
	$\bm{s}^{(1)}_t$ & Vector representing the state of the devices at time $t$.\\
	$\bm{s}^{(2)}_t$ & Vector representing the state of the modulation instructions sent to controllable devices, at time $t$.\\
	$\bm{flex}_t$ & Vector representing the state of the flexible loads at time $t$, it is a part of $\bm{s}^{(2)}_t$.\\
	$\bm{s}^{(3)}_t$ & Part of the state of the system that, at time $t$, keeps track of past realizations of the uncertain
	phenomena.\\
	$\bm{s}_t$ & Global state of the system at time $t$.\\
	$\bm{a}_t$ & Vector of control actions taken at time $t$.\\
	$\bm{\overline{p}}_t$ & Maximum level of active power injection for period $t+1$ and for each of the generators $g \in \mathcal{G}$, part of $\bm{a}_t$.\\
	$\bm{act}_t$ & Activation indicators of the flexibility services of the loads $d\in \mathcal{F}$, part of $\bm{a}_t$.\\
\end{tabularx}
\noindent\begin{tabularx}{\textwidth}{r p{10cm}}
	$act_{d,t}$ & Component of $\bm{act}_t$ for flexible load $d \in {\mathcal{F}}$.\\
	$\bm{w}_t $ & Information on exogenous phenomena available at time $t$.\\
\end{tabularx}

\paragraph{Operators and functions:} {~} \\
\begin{tabularx}{\textwidth}{r p{10cm}}
	$|\cdot|$ & Magnitude of a complex number or size of a set.\\
	${\cdot}^*$ & Complex conjugate.\\
	$f: \mathcal{S} \times \mathcal{A}_{\bm{s}} \times \mathcal{W} \rightarrow \mathcal{S}$ & Transition function of the system.\\ 
	$r : \mathcal{S} \times \mathcal{A}_{\bm{s}} \times \mathcal{S} \to \mathbb{R}$ & Reward function.\\
	$\pi : \mathcal{S} \to \mathcal{A}_{\bm{s}} $ & Policy that returns an action for every feasible state.\\
	$\mathcal{G}(n)$& Set of generators connected to node $n$.\\
	$\mathcal{C}(n)$& Set of loads connected to node $n$.\\
	$\mathcal{F}(n)$& Set of flexible loads connected to node $n$.\\
\end{tabularx}

\section{Introduction}

In Europe, the 20/20/20 objectives of the European Commission and the consequent financial incentives established by local governments are currently driving the growth of electricity generation from renewable energy sources \cite{res_support}. A substantial part of the investments is made at the distribution networks (DN) level and consists of the installation of wind turbines or photovoltaic panels. The significant increase of the number of these distributed generators (DGs) undermines the \emph{fit and forget}\footnote{Under this approach, adequate investments in network components (i.e., lines, cables, transformers, etc.) must be made in order to always avoid congestion and voltage problems.} doctrine, which has dominated the planning and the operation of DNs up to now. This doctrine was developed when the energy was coming from the transmission network (TN) to the consumers, through the distribution network (DN). With this approach, adequate investments in network components (i.e., lines, cables, transformers, etc.) are made to avoid congestion and voltage issues, without requiring continuous monitoring and control of the power flows or voltages. To that end, network planning is done with respect to a set of critical scenarios gathering information about production and demand levels, in order to always ensure sufficient operational margins. Nevertheless, with the rapid growth of DGs, the preservation of such conservative margins implies significant network reinforcement costs\footnote{Network reinforcement is the process of upgrading the transmission capacity of lines, cables, transformers, and other devices. As distribution systems of interest in this paper are mostly done of underground cables, upgrading them implies a lot of infrastructure work.}, because the net energy flow may be reversed, from the distribution network to the transmission network, and flows within the distribution network be very different from the flows historically observed.

In order to host a larger share of distributed generation \cite{cornelusse2015global} and avoid potentially prohibitive reinforcement costs \cite{wang2010dg}, \emph{active network management} (ANM) strategies have recently been proposed as alternatives to the \textit{fit and forget} approach. The principle of ANM is to address congestion and voltage issues via short-term decision-making policies \cite{joseph_et_al}. Frequently, ANM schemes maintain the system within operational limits in quasi real-time by relying on the curtailment of wind or solar generation \cite{liew_goran, nando_capacity, dolan}. Curtailment of renewable energy may, however, be very controversial from an environmental point of view and should probably be considered as a last resort. In that mindset, it is worth investigating ANM schemes that could also exploit the flexibility of the loads, so as to decrease the reliance on generation curtailment. Exploiting flexible loads within an ANM scheme comes with several challenges. One such challenge is that modulating a flexible load at one instant is often going to influence its modulation range at subsequent instants. This is because flexible loads (e.g. heat pumps) are often constrained to consume a specific amount of energy over a certain duration. In this context, it is therefore important for a distribution system operator (DSO) to take decisions by planning operations over a sufficiently long time horizon \cite{gemine2013active, gill2014dynamic, macedo2015optimal}. The uncertainty of future power injections from DGs relying on natural energy sources (wind, sun, \textit{etc.}), as well as the uncertainty of the power consumption of the loads, should also be explicitly accounted for in the ANM strategy. In this work we consider the operation of the medium-voltage (MV) network of the DSO, i.e. low voltage subnetworks are aggregated, since in general current DSOs' dispatching centers only monitor the MV part, and ANM in low voltage distribution systems is generally performed using distributed algorithms \cite{olivier2015active}. 

Many authors have already attempted to provide solutions to these operational planning problems. Since they rely on different formulations, it is difficult for one author to rebuild on top of another's work. However, these formulations can be considered as an extension of the \emph{optimal power flow} (OPF) problem \cite{dommel1968optimal}. More specifically, they can be assimilated to sequential decision-making problems where, at each time step, constraints that are similar to those used for defining an OPF problem are met. 
Optimal power flow problems, although non-convex, have been solved for a long time using local nonlinear optimization methods. Interior-point methods are probably the most widespread class of methods dedicated to this problem \cite{capitanescu2007ipm}. If the solution they provide has no guarantee to be globally optimal, then they have been made popular by their convergence speed and their ability to solve problems of large dimensions fairly efficiently. Convexifications of the power flow equations have been successful, in particular in \cite{jabr2006radial} where the author models power flows in a radial distribution system using second-order cone constraints. Recently, semidefinite programming (SDP) was applied as a convex relaxation to the OPF problem \cite{lavaei2012}. The authors report no duality gap on some standard meshed test systems and randomized versions of these test systems. The zero duality gap property was thus observed experimentally on standard test systems, and further research resulted in sufficient conditions. This is the case, for example, if the objective function is convex and monotonically increasing with the active power generation, and if the network has a radial topology \cite{bose2012quadratically, gan2012branch}.  Another approach aiming at global optimality relies on Lagrangian relaxation (LR) \cite{phan2012}. The author also describes a spatial branch and bound (B\&B) algorithm to close the gap, should one exist. The ability of both SDP and LR to decrease the optimality gap within a B\&B framework was evaluated in \cite{gopalakrishnan2012global}. Although SDP appeared to be computationally more attractive, it showed that it could be very challenging to reach a significant gap reduction within reasonable time limits, even for small test systems. A different approach is considered in \cite{bolognani2016existence}, where the authors present a linear approximation of the power flow equations with a focus on distribution networks.
Multi-period applications related to energy storage are investigated in \cite{gayme2011optimal}, where the SDP relaxation of \cite{lavaei2012} is successfully applied, as their particular application met the conditions of having no duality gap. The authors of \cite{gopalakrishnan2013global} argue that extending \cite{gopalakrishnan2012global} to a multi-period setting yields an SDP too large for current solvers to solve efficiently and suggest relaxing the time-coupling constraints using LR.  However, it ended up being computationally too expensive to make the B\&B approach worthwhile. Many papers consider the unit commitment problem over an AC network, which is an instance of a multi-period OPF with discrete variables. For instance in \cite{alguacil2000multiperiod}, a generalized Benders decomposition divides the problem into a linear master problem with discrete variables and nonlinear multi-period sub-problems. Benders cuts are generated from the sub-problems to tighten the MIP master problem. Finally, \cite{gemine2014relaxations} focused on trying to solve a problem that is mathematically close to the one we consider and provides more information on related research. 

A first objective of this work is to facilitate the comparison of solution techniques that have been developed in the research community. To that end, we first propose a generic formulation of ANM related decision-making problems. More specifically, we detail a procedure to state these problems as Markov Decision Processes (MDP), where the system dynamics describes the evolution of the electrical network and devices, while the action space encompasses the control actions that are available to the DSO. Afterwards, we instantiate this procedure on networks of 5, 33, and 77 buses, and use the elements of the resulting MDPs to build a simulator of these systems, which is available at \url{http://www.montefiore.ulg.ac.be/~anm/}. 
As a second contribution, we provide quantitative results for the resolution of the ANM problem cast as a stochastic mixed-integer nonlinear program (MINLP), as well as a mixed-integer second-order cone programming (MISOCP) relaxation and a mixed-integer linear programming (MILP) approximation, using state of the art open source and commercial solvers. We then perform a sensitivity analysis over the size of the distribution system, the amount of flexibility available in the system, and the number of scenarios considered in the deterministic equivalent of the stochastic program.
Finally, a last contribution lies in the features modeled in this work. Compared to the work of \cite{gill2014dynamic} and \cite{macedo2015optimal}, we explicitly account for uncertainty, and for discrete variables stemming from the activation of flexibility services. Compared to our work, \cite{gill2014dynamic} relies only on a continuous nonlinear programming formulation, and thus does not analyze linear or second order cone programming formulations, but models a storage system, and  \cite{macedo2015optimal} also models discrete decisions variables, but they are related to capacitor banks switching and storage system operation modes. The latter reference also uses MISOCP and MILP formulations.

The rest of this paper is structured as follows. The ANM problem of a DN is described in Section~\ref{sec:probStatement}, where the electrical model and the network operation details are explained, and the operational planning problem is formulated as a Markov decision process. This formulation is then cast as a stochastic mixed-integer nonlinear program in Section~\ref{sec:solution}, where a second order cone relaxation and a linear approximation are also detailed.
The test beds built around the different distribution systems are described in Section~\ref{sec:test}, and test results are presented in Section~\ref{sec:num_results}.
Section~\ref{sec:conclusion} concludes and presents possible extensions of this work.

\section{Problem Description}
\label{sec:probStatement}

\subsection{Model of the electrical distribution system}
\label{EDSmodel}

In this paper, we are always considering that the network and all its devices are operating in alternating current mode. We also make the choice to represent complex numbers in rectangular coordinates.

\begin{figure}[tb]
\centering
\scalebox{0.8}{\begin{circuitikz}[american voltages]
    \large
	\draw (0,2.1) node[transformer](T){} (T.base) node[above]{1:$t_{mn}$} (2,2.1);
	\draw (-1,2.1) node[above]{$V_m$} to [short, *-] (-0.5,2.1);
	\draw (1,2.1) to (2,2.1);
	\draw (5,2.1) to (5.5, 2.1);
	\draw (2,2.1) to [R, l_=$Y_{mn}^{(br)}$,i=$I_{mn}$] (5,2.1);
	\draw (5.5,2.1) to (6,2.1);	
	\draw (7.5,2.1) to [short, -*] (8,2.1) node[above]{$V_n$};
	\draw (2,0) to [R, l_=$Y_{mn}^{(sh)}$] (2,2.1);
	\draw (5,0) to [R, l^=$Y_{nm}^{(sh)}$] (5,2.1);
	\draw (7,2.1) node[transformer](T2){} (T2.base) node[above]{$t_{nm}$:1} (8.5,2.1);	
	\draw (1,0) to (6.5,0);
	\draw(-1,0)  to [short, *-] (-.5,0);
	\draw (7.5,0)  to [short, -*] (8,0);
\end{circuitikz}}
\caption{$\pi$-model of a link.}
\label{fig:pi_model}
\end{figure}
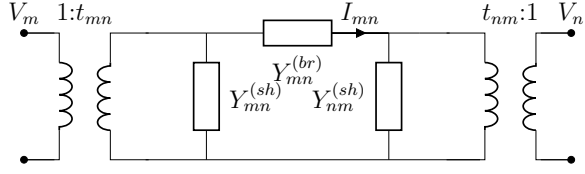

The electrical distribution system can be mathematically represented by a graph, that is a set of nodes, and a set of links connecting nodes. 
A node is an electrical bus characterized by a voltage $V_{n} \in \mathbb{C}$.
In addition to links connecting a bus to its neighbors, several devices may be connected to a bus. Devices are either injecting or withdrawing power. 
Every link $(m,n) \in \mathcal{L} \subset \mathcal{N}^2$ connects a pair of nodes $m, n \in \mathcal{N}$ and represents an overhead line, an underground cable, or a transformer.
A link is represented by its $\pi$-model, composed of five complex parameters: two ratios $t_{mn}$ and $t_{nm}$, a branch admittance $Y^{(br)}_{mn}$, and two shunt admittances $Y^{(sh)}_{mn}$ and $Y^{(sh)}_{nm}$ (see Fig.~\ref{fig:pi_model}), that are considered fixed in this work, although opportunities to change them dynamically can exist in practice. More details on the $\pi$-model of specific links can be found in \cite{andersson2004modelling}.

To ensure the proper operation of  the devices connected to a bus, the voltage magnitude $|V_{n}|$ at note $n$ should not deviate too much from its nominal voltage level: 
\begin{align}
\forall n \in \mathcal{N} : \underline{V}_n \leq |V_{n}| \leq \overline{V}_n\,.\label{eq:v_limits}
\end{align}
If $I_{mn} \in \mathbb{C}$ is the branch current through link $(m,n)$, its magnitude $|I_{mn}|$ should be kept below a pre-specified limit to prevent excessive heating of conductors and insulating materials:
\begin{align}
\forall (m,n) \in \mathcal{L} : |I_{mn}| \leq \overline{I}_{mn}.\label{eq:i_max}
\end{align}
In reality, there are several limits depending on the magnitude and duration of the over-current. In this work we consider only the most conservative limit, since we want to keep a sufficient margin as we are taking decisions ahead of time with a relatively high uncertainty.
The magnitude of the current $I_{l}$ in link $l$ connecting nodes $m$ and $n$ can be deduced from the voltage at these nodes by
\begin{align}
|I_{mn}| = \left|\left(|t_{mn}|^2 V_{m} -(t^{(l)}_{mn})^* t^{(l)}_{nm} V_{n}\right)Y^{(br)}_{mn}\right|\,,\label{eq:branch_current}
\end{align}
where $\cdot^*$ denotes the complex conjugate operator.

Before defining the power injections as a function of voltages, it is convenient to relate the current injected at nodes to the voltage by writing:
\begin{align}
\bm{I} = \bm{Y}\bm{V}\,,\label{eq:current}
\end{align}
where $\bm{I} = (I_1, \dots, I_{|\mathcal{N}|})$ is the vector of the current injection at nodes, $\bm{V} = (V_1, \dots, V_{|\mathcal{N}|})$ is the vector of the voltage at nodes, and $\bm{Y}$ is the $|\mathcal{N}| \times |\mathcal{N}|$ nodal admittance matrix, which has its elements defined by
\begin{align}
Y_{mn} = \begin{cases}
-(t^{(l)}_{mn})^* t^{(l)}_{nm} Y^{(br)}_{mn}&\text{if}~m \neq n~\text{and}~\exists (m,n) \in \mathcal{L}\,,\\
\sum_{(m,k) \in \mathcal{L}} |t_{mk}|^2 (Y^{(sh)}_{mn}+Y^{(br)}_{mn})&\text{if}~m = n\,,\\
0&\text{otherwise}\,.
\end{cases}
\end{align}
Regarding the active power $P_n$ and reactive power $Q_n$ injected at every node $n$, they are related to the node voltages through the power flow equations \cite{monticelli1999state}:
\begin{align}
\forall n \in \mathcal{N} : S_n = P_n + jQ_n = V_n I^*_n = V_n \bm{Y}^*_{n\cdot} \bm{V}^*\,,\label{eq:pf_eqs}
\end{align}
where $S_n$ is the apparent power injection at bus $n$ and $\bm{Y}_{n\cdot}$ denotes the $n^{\text{th}}$ row of the nodal admittance matrix.
By convention a power injection is positive if it supplies the network and negative if it takes energy from the network.

In summary, there are four quantities attached to each node $n \in \mathcal{N}$ that determine the electrical state of the system: $P_n$,  $Q_n$, and real and imaginary parts of $V_n$. The power flow equations (\ref{eq:pf_eqs}) provide $2|\mathcal{N}|$ relations. $2|\mathcal{N}|$ variables should thus be fixed to obtain a solution to this system of equations. In general, $V_n$ is fixed on one side of the transformer between the MV network and the transmission system, to provide a reference voltage.  At other nodes, the active power injection $P_n$ is known, as well as either the reactive power $Q_n$ or the voltage magnitude $|V_n|$, depending on the type of device connected at the node.  In this work we consider that we have some control over the power flows in the system, hence we consider that less than $2|\mathcal{N}|$ variables are fixed and that we can act on $P_n$ and $Q_n$ at some nodes. The system is actually controlled by acting on the electrical devices attached to these nodes.  

\begin{figure}[t]
\centering
\includegraphics[width=5in]{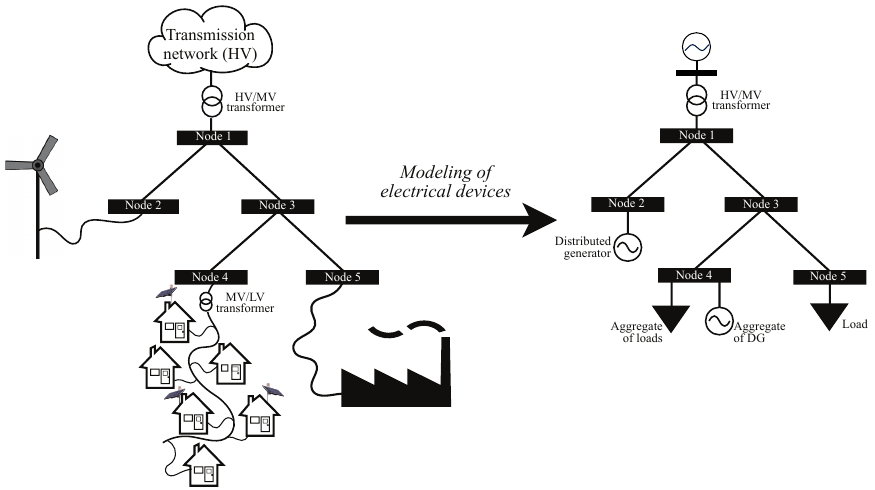}
\caption{System model}
\label{fig:devices}
\end{figure}

Electrical devices can be classified into two distinct subsets,  the set $\mathcal{C} \subset \mathcal{D}$ of loads  that withdraw power from the network, and 
the set $\mathcal{G} \subset \mathcal{D}$ of generators that inject power into the network.
Within each subset, we also distinguish two types of device models. The first ones represent individual injection and withdrawal points. They can model certain types of DGs or consumers that are directly connected to the MV grid (e.g., wind farms, some companies and factories, etc.). The others model an aggregate set of devices that are assimilated to a single connection point at the MV grid (e.g., residential consumers and solar panels). Correspondences between some physical elements and their device model are illustrated in Fig.~\ref{fig:devices}. At node 3, a set of residential loads and a set of distributed solar units have been aggregated into a single load model and a single generator model.

\begin{figure}[hbt] 
	\begin{tikzpicture}[scale=0.5,
	thick,
	>=stealth',
	dot/.style = {
		draw,
		fill = white,
		circle,
		inner sep = 0pt,
		minimum size = 4pt
	}
	]
	\coordinate (O) at (0,0);
	\draw[line width=0.75mm, fill=gray!20] (0,4) -- (5,4) -- (7,2) -- (7,-2) -- (5,-4) -- (0,-4) -- (0,4);
	\draw[->] (-0.3,0) -- (8,0) coordinate[label = {below:$P_g$}] (xmax);
	\draw[->] (0,-5) -- (0,5) coordinate[label = {right:$Q_g$}] (ymax);
	\draw (0,-5) -- (0,5);
	\draw (7,-5) -- (7,5) node[right] {$P_{max,g}$};	
	\draw (0,4) -- (8,4) node[right] {$Q_{max,g}$};
	\draw (0,-4) -- (8,-4) node[right] {$Q_{min,g}$};
	\draw (4,5) node[above] {$Q_g \leq \alpha^+_g P_g + \beta^+_g$} -- (8,1);
	\draw (4,-5) node[below] {$Q_g \leq \alpha^-_g P_g + \beta^-_g$} -- (8,-1);
	\end{tikzpicture}
	\caption{Illustration of a polyhedral set $\mathcal{O}_d$ defining the P-Q capability area of a distributed generator $g \in \mathcal{G}$.}
	\label{fig:P-Q}
\end{figure}
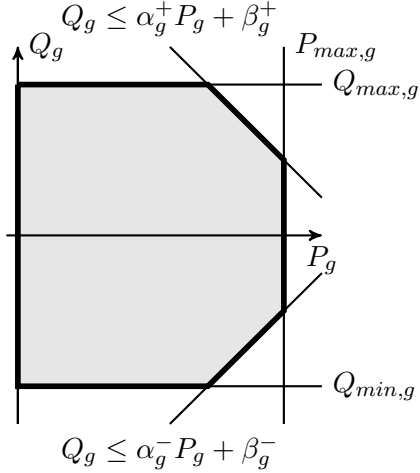

An active power injection value $P_{d}$ and a reactive power injection value $Q_{d}$ are associated with every device $d \in \mathcal{D}$, and, denoting the set of devices connected at node $n$ by $\mathcal{D}(n) \subset \mathcal{D}$:
\begin{align}
\forall n \in \mathcal{N} : S_{n} &= P_n + j Q_n = \sum_{d \in \mathcal{D}(n)} (P_{d} + j Q_{d})\,.\label{eq:s_n}
\end{align}
Every device $d$ has a restricted set $\mathcal{O}_d \subset \mathbb{R}^2$ of valid ($P_d$,$Q_d$) injection points. We assume that the loads are operating at a constant power factor, i.e. the ratio between reactive and active powers - denoted as $\tan{\phi_d}$ - remains unchanged:
\begin{align}
\forall d \in   \mathcal{C} : \mathcal{O}_d = \{ (P_d,Q_d) \in \mathbb{R}^2~|~\frac{Q_{d}}{P_{d}} = \tan{\phi_d}\}\,.\label{eq:load_pf}
\end{align}
For distributed generators, the injections points have to stay within a polyhedron, as illustrated in Fig. \ref{fig:P-Q}. This set is defined by lower and upper bounds on both $P_d$ and $Q_d$, as well as by two linear constraints that prevent a full flexibility on $Q_g$ when $P_g$ is close to its maximum. These constraints model the limitations of the power converter and/or of the electric generator \cite{engelhardt2011reactive}. We have:
\begin{align}
\forall g \in \mathcal{G} : \mathcal{O}_g = \{ (P_d,Q_d) \in \mathbb{R}^2~|~&P_{min,g} \leq P_g \leq P_{max,g}\,,\nonumber\\
&Q_{min,g} \leq Q_g \leq Q_{max,g}\,,\nonumber\\
&Q_g \leq \alpha^+_g P_g + \beta^+_g\,,\nonumber\\
&Q_g \leq \alpha^-_g P_g + \beta^-_g\}\,.\label{eq:valid_pq}
\end{align}


\subsection{Operational planning problem statement}
\label{sec:operationalPlanning}
Considering the model of the electrical distribution network presented in Section~\ref{EDSmodel}, operational planning is a recurring task performed by the DSO to anticipate the evolution of the system, that is the impact of the evolution of the injection and the consumption patterns on the operational limits of the system, and take preventive decisions to stay within these limits. Among the available decisions in the considered timing of operations, we consider that acting on the power injected or consumed by a predefined set of devices is the only type of control the DSO has, as detailed in Section~\ref{EDSmodel}.
We describe the evolution of the system by a discrete-time process having a time horizon $\mathcal{T}$, the number of periods used for the operational planning phase. 
The period duration is 15 minutes,  by analogy with the typical duration of a market period.  
The power injection and withdrawal levels are constant within a single period, and we neglect the fast dynamics of the system, which may be handled by real time controllers \cite{soleimani2016receding}.
The control actions in this section are aimed to directly impact these power levels and can introduce time-coupling effects, depending on the type of device.
We now describe two control means of the system, the modulation of the generation and the modulation of the demand, as well as  one of  the possible interaction  schemes between  the actors  of this  system. 

\begin{figure}[tb] 
\centering
\includegraphics[width=3.25in]{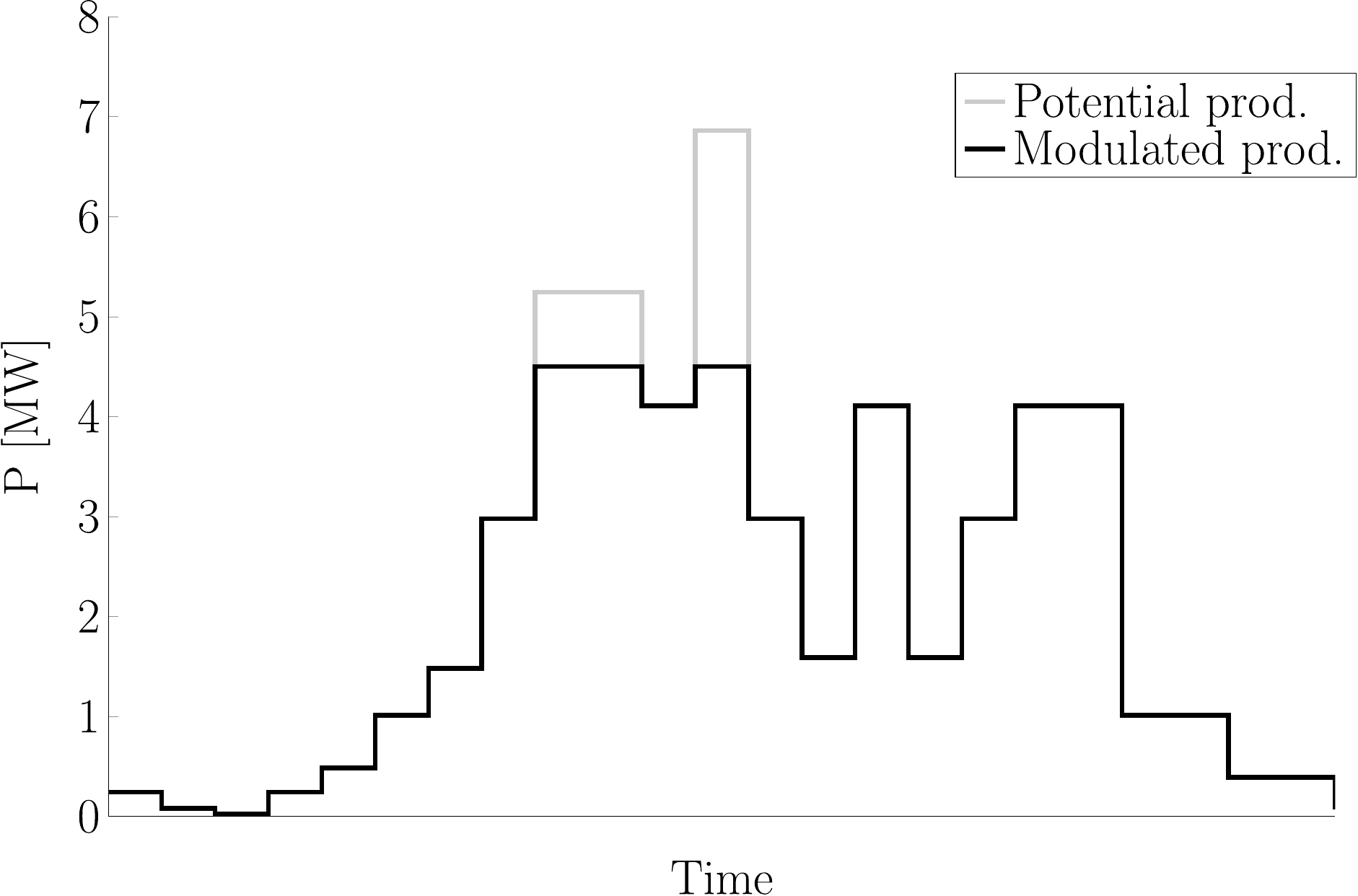}
\caption{Curtailment of a distributed generator.}
\label{fig:curt_ex}
\end{figure}

For each device belonging to the set $\mathcal{G} \subset \mathcal{D}$ of DGs, the DSO can impose a curtailment instruction, i.e.\ an upper limit on the generation level of the DG (cf. Fig.~\ref{fig:curt_ex}).
This request can be performed until the time period immediately preceding the one concerned by the curtailment and it is acquired in exchange for a fee. 
This fee is used to compensate the producer for financial loss related to the energy that could not be produced during modulation periods. We assume that this fee is defined as a per unit compensation for the energy not produced, with respect to the actual potential that is known after the market period.

\begin{figure}[tb]
\centering
\subfloat[Modulation signal of the consumption ($T_d=9$).\label{fig:flex_ex_1}]{\includegraphics[width=3.25in]{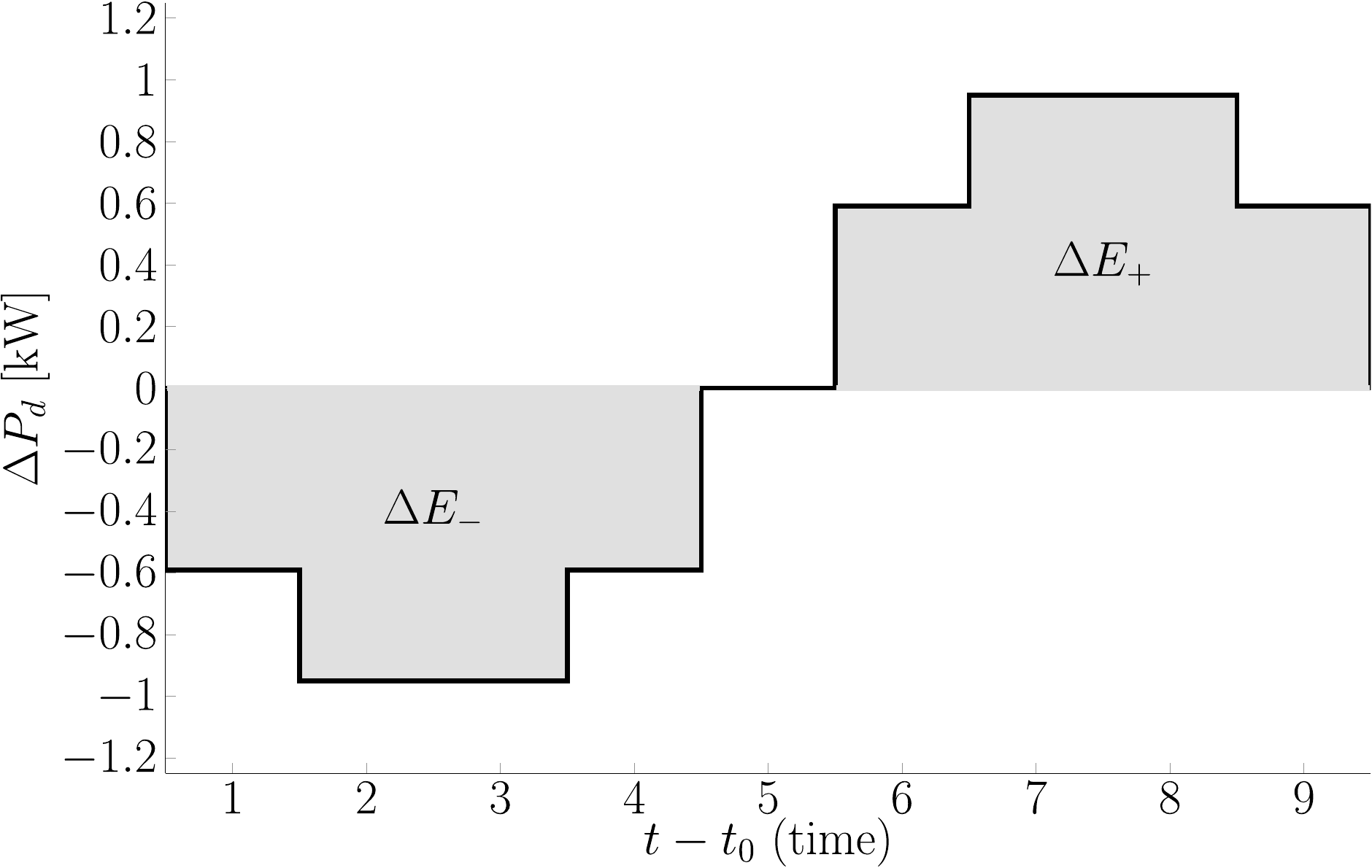}}\\
\subfloat[Impact of the modulation signal over the consumption. \label{fig:flex_ex_2}]{\includegraphics[width=3.25in]{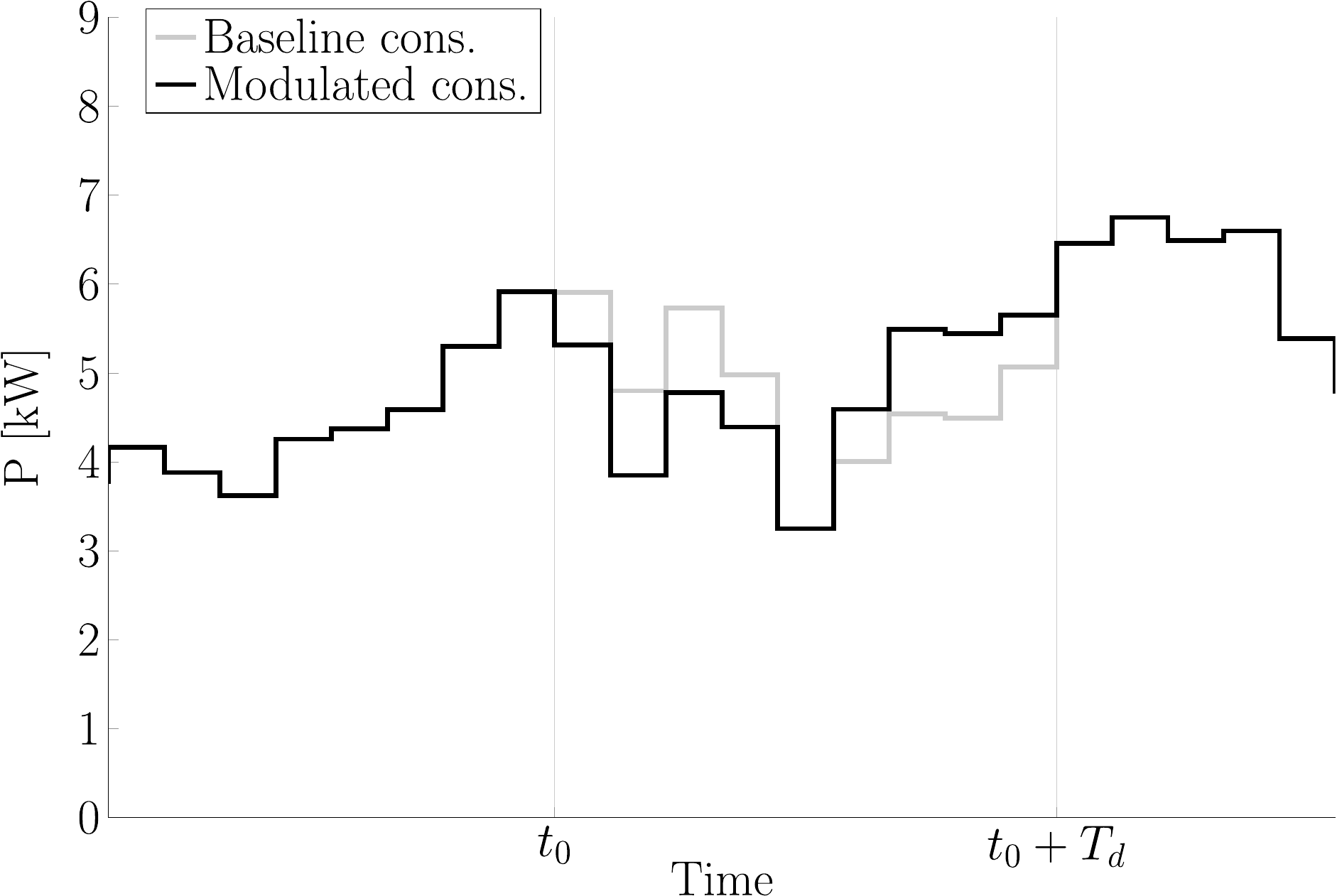}}\\
\caption{Illustration of flexibility services.\label{fig:flex_ex}}
\end{figure}

We also consider that the DSO can modify the consumption of some flexible loads, a subset $\mathcal{F}$ of full set of the loads $\mathcal{C} \subset \mathcal{D}$ of the network.
An activation fee is associated with this control mean and flexible loads can be notified of activation until the time immediately preceding the start of the service. 
Once the activation is performed at time $t_0$, the consumption of the flexible load $d$ is modified by a certain value during $T_d$ periods. 
For each of these modulation periods $t \in \{t_0+1,..., t_0 + T_d\}$, this value is defined by the modulation function $\Delta P_{d}(t - t_0)$. 
An example of modulation function and its influence over the consumption curve are presented in Fig.~\ref{fig:flex_ex}.
Loads cannot be modulated in an arbitrary way. There are constraints to be imposed on the modulation signal, which are inherited from the flexibility sources of the loads, such as an inner storage capacity (e.g. electric heater, refrigerator, water pump) or a process that can be scheduled with some flexibility  (e.g., industrial production line, dishwasher, washing machine). In any case, we will always consider that the modulation signal $\Delta P_d$ has to satisfy the following conditions:
\begin{itemize}
\item A downward modulation is followed by an increase of the consumption, and conversely.
\item The integral of the modulation signal is null in order to ensure that the consumption is only shifted, not modified.
\end{itemize}

Other approaches that we do not consider in this work exist to control the system, such as modulating the tariff signal(s), acting on the topology of the network, or using distributed storage sources. We do not model either the automatic regulation devices that often exist in distribution systems, such as On Load Tap Changers of transformers which automatically adapt to control the voltage level. This should be the case, obviously, in a real life application. We will discuss in the conclusion what are the implications of these non-modeled control possibilities.

\subsection{Optimal sequential decision-making formulation}
\label{sec:dynamics}
We now formulate operational planning as an optimal sequential decision-making problem, that is, we explain how the time, the decision process, and uncertainty are included to extend the mathematical model described in Section~\ref{EDSmodel}.
The sequential aspect is induced by the modulation service that is provided by flexible loads. Indeed, if such a service is activated at time $t_0$ for a flexible load $d$, the action will influence the system for the set of periods $\{t_0 + 1,..., t_0+T_d\}$. In addition to being a sequential problem, it is also stochastic, because the evolution of the system and the outcome of control actions are affected by several uncertain factors. These factors include, but are not limited to, the wind speed, the level of solar irradiance, and the consumption level of the loads. In this section, we model this problem as a Markov decision process with mixed-integer sets of states and actions. We thus consider that the transition probabilities of the state of the system from a period $t$ to a period $t+1$ only depend on the state at time $t$. However, this state can encompass several past values of wind speed, solar irradiance, consumption levels, and any auxiliary modeling variables, in order to obtain a relevant model. An automatic procedure that determines an adequate number of past values to track in the state is presented in Section~\ref{sec:test}. Note that modeling the actual system as a Markovian system is not restrictive as all properly modeled systems are Markovian if the state variables capture all the information to model the system from time $t$ and onwards \cite{powell2016unified}. Finally, the notion of optimality is defined using a reward function that associates an immediate reward (or score) to every transition of the system. The better the cumulated reward over a system trajectory, the better the sequence of control actions for this trajectory.

\subsubsection{System state}

The global state space $\mathcal{S}$ of the system is decomposed in three subsets:
\[
\mathcal{S} = \mathcal{S}^{(1)} \times \mathcal{S}^{(2)} \times \mathcal{S}^{(3)}.
\]

The power injections of the devices are sufficient to obtain the value of the electrical quantities through equations~(\ref{eq:pf_eqs}) and (\ref{eq:s_n}). These injections are determined from the realization of the exogenous consumption and generation processes at a given time period, and from the modulation instructions for that period. If the consumption processes require the representation of the individual consumption of every load, it is possible to obtain the production of DGs given the power level of their energy source (i.e. the wind speed or the level of solar irradiance). We thus define a first state set $\mathcal{S}^{(1)}$ such that the vectors $\bm{s}^{(1)}_t \in \mathcal{S}^{(1)}$ are defined by
\[
\bm{s}^{(1)}_t = (P_{1,t}, \dots, P_{|\mathcal{C}|,t}, ir_{t}, v_{t})\,,
\]
where, at time $t \in \mathcal{T}$, the $ir_t$ and $v_t$ components represent the level of solar irradiance and the wind speed, respectively. If, for the sake of simplicity, we consider only solar and wind generation, other types of generators could easily be integrated by increasing the dimension of $\mathcal{S}^{(1)}$. Note that the reactive power withdrawals of loads are known from $\bm{s}^{(1)}_t$ through equation~(\ref{eq:load_pf}). 

The vector $\bm{s}_t^{(2)} \in \mathcal{S}^{(2)}$, defined as
\[
\bm{s}^{(2)}_t = (\overline{P}_{1,t}, \hat{Q}_{1,t}, \dots, \overline{P}_{|\mathcal{G}|,t}, \hat{Q}_{|\mathcal{G}|,t}, flex_{1,t}, \dots, flex_{|\mathcal{F}|,t}),
\]
contains the upper limits $\overline{P}_{g,t}$ on the active power injection and the reactive set-points $\hat{Q}_{g,t}$ of the DGs $g \in \mathcal{G}$, as authorized by the DSO, and the indicators $flex_{d,t}$ of the flexibility service state of the loads $d \in \mathcal{F}$:
\[
 flex_{d,t} =
  \begin{cases}
   \text{number of active periods left}	& \text{if service is active}\\
   0       			& \text{if service is inactive}\,.
  \end{cases}
\]

We denote by $\bm{s}^{(3)}_t \in \mathcal{S}^{(3)}$ the part of the system's state that, at time $t \in \mathcal{T}$, keeps track of past realizations of the uncertain phenomena (i.e. wind speed, solar irradiance, and consumption levels) and contains the optional auxiliary modeling variables. Its purpose is to improve the accuracy of the stochastic modeling and to allow the representation of processes that are required for some reward functions (see Section~\ref{sec:reward}). The number of past values can be different for each phenomenon and, we have
\begin{align}
\bm{s}^{(3)}_t = (&P_{1,t-1}, \dots, P_{1,t-N_{loads}+1}, \dots,  P_{|\mathcal{C}|,t-1}, \dots, P_{|\mathcal{C}|,t-N_{loads}+1}, \notag \\
&ir_{t-1}, \dots, ir_{t-N_{ir}+1}, v_{t-1}, \dots, v_{t-N_{v}+1}, \notag \\
&s^{(\mathrm{aux})}_{1,t}, \dots, s^{(\mathrm{aux})}_{N_{aux},t}) \notag
\end{align}
where $N_{loads}$, $N_{ir}$, $N_{v} \in \mathbb{Z}_0^{+}$, and $N_{aux} \in \mathbb{Z}^{+}$. The value of these parameters has to be determined when instantiating the presented abstract decision model (see Section~\ref{sec:test}). A value of $1$ for the three former parameters means that the history of the corresponding phenomenon consists of $\bm{s}^{(1)}$ only, while a value of $0$ for the latter parameter means that there is no auxiliary variable. We denote thereafter the vector of the $N_{aux}$ auxiliary modeling variables by $\bm{s}^{\mathrm{(aux)}}_t$.

\subsubsection{Control actions}
\label{sec:control_actions}
The control means that are available to the DSO to control the system are modeled by the set $\mathcal{A}_{\bm{s}}$ of control actions. This set depends on the state $\bm{s}_t$ of the system because it is not possible to activate the flexibility service of a load if it is already active. The components of vectors $\bm{a}_t \in \mathcal{A}_{\bm{s}}$ are defined by
\[
\bm{a}_t = (\bm{\overline{\bm{p}}}_t, \hat{\bm{q}}_t, \bm{act}_{t})\,,
\]
with $\overline{\bm{p}}_t, \hat{\bm{q}}_t \in \mathbb{R}^{|\mathcal{G}|}$ such that, for period $t+1$ and for each of the generators $g \in \mathcal{G}$, $\overline{p}_{g,t}$ and $\hat{q}_{g,t}$ indicate the maximum level of active power injection and the desired reactive set-point, respectively. On the other hand, the vector $\bm{act}_{t}$ represents the activation indicators of the flexibility services of the loads $d\in \mathcal{F}$, where each component $act_{d,t}$ belongs to $\mathcal{A}_{d,\bm{s}}$, which is defined as
\begin{align}
 \mathcal{A}_{d,\bm{s}} =
  \begin{cases}
   \{0,1\}	& \text{if } flex_{d,t} = 0\\
   \{0\}       	& \text{if } flex_{d,t} > 0\,,
  \end{cases}\label{eq:feas_actions}
\end{align}
to ensure that a load which is already active is not activated.

By using this representation of the control actions, we consider that a curtailment or flexibility activation action targeting a period $t$ must always be performed at the period $t-1$, as described in Section~\ref{sec:probStatement}. We do not consider the possibility to notify control actions several periods ahead, because it would induce even larger time-coupling effects, while not improving the extent of control of the DSO since in the interaction model considered in this paper the cost associated with an action is independent of the notification delay.

\subsubsection{Transition function}

The system evolution from a state $\bm{s}_t$ to a state $\bm{s}_{t+1}$ is described by the transition function $f$. The new state $\bm{s}_{t+1}$ depends, in addition to the preceding state, on the control actions $\bm{a}_t$ and on the realization of the stochastic processes:
\[
f : \mathcal{S} \times \mathcal{A}_{\bm{s}} \times \mathcal{W} \to \mathcal{S}\,,
\]
where $\mathcal{W}$ is the set of possible realizations of a random process. The general evolution of the system is thus governed by relation
\begin{align}
\bm{s}_{t+1} = f(\bm{s}_t, \bm{a}_t, \bm{w}_t)\,,\label{eq:trans_fct}
\end{align}
where $\bm{w}_t \in \mathcal{W}$ represents the exogenous information and follows a probability law $p_{\mathcal{W}}(\cdot)$. We could write equivalently that $\bm{s}_{t+1} \sim p_{\mathcal{S}}(\cdot | \bm{s}_t, \bm{a}_t)$, which clearly highlights that the next state of the system follows a probability distribution that is conditional on the current state and on the action taken at the corresponding time step. However, we favor notation of equation~(\ref{eq:trans_fct}) as it enables an easier formulation of concepts that are introduced latter in this paper. We now describe the various elements that constitute the transition function.

\paragraph{Load consumption}
The uncertainty about the behavior of consumers inevitably leads to uncertainty about the power level they draw from the network. However, over a one-day horizon, some trends can be observed. For example, consumption peaks arise in the early morning and in the evening for residential consumers, but at levels that fluctuate from one day to another and among consumers. We model the evolution of the consumption of each load $d \in \mathcal{C}$ by
\begin{align}
P_{d,t+1} = f_d(P_{d,t}, P_{d,t-1}, \dots, P_{d,t-N_{loads}+1}, \bm{s}^{(aux)}_t, \bm{w}_{d,t})\,,\label{eq:trans_cons}
\end{align}
where $\bm{w}_{d,t} \sim p_{\mathcal{W}_d}(\cdot)$ denotes some components of $\bm{w}_t \sim p_{\mathcal{W}}(\cdot)$. Given the hypothesis of a constant power factor for the loads, the reactive power consumption can directly be deduced from $P_{d,t+1}$:
\begin{align}
Q_{q,t+1} = \tan{\phi_d} \cdot P_{d,t+1}\,.
\end{align}

\begin{figure}[t] 
\centering
\includegraphics[width=3.25in]{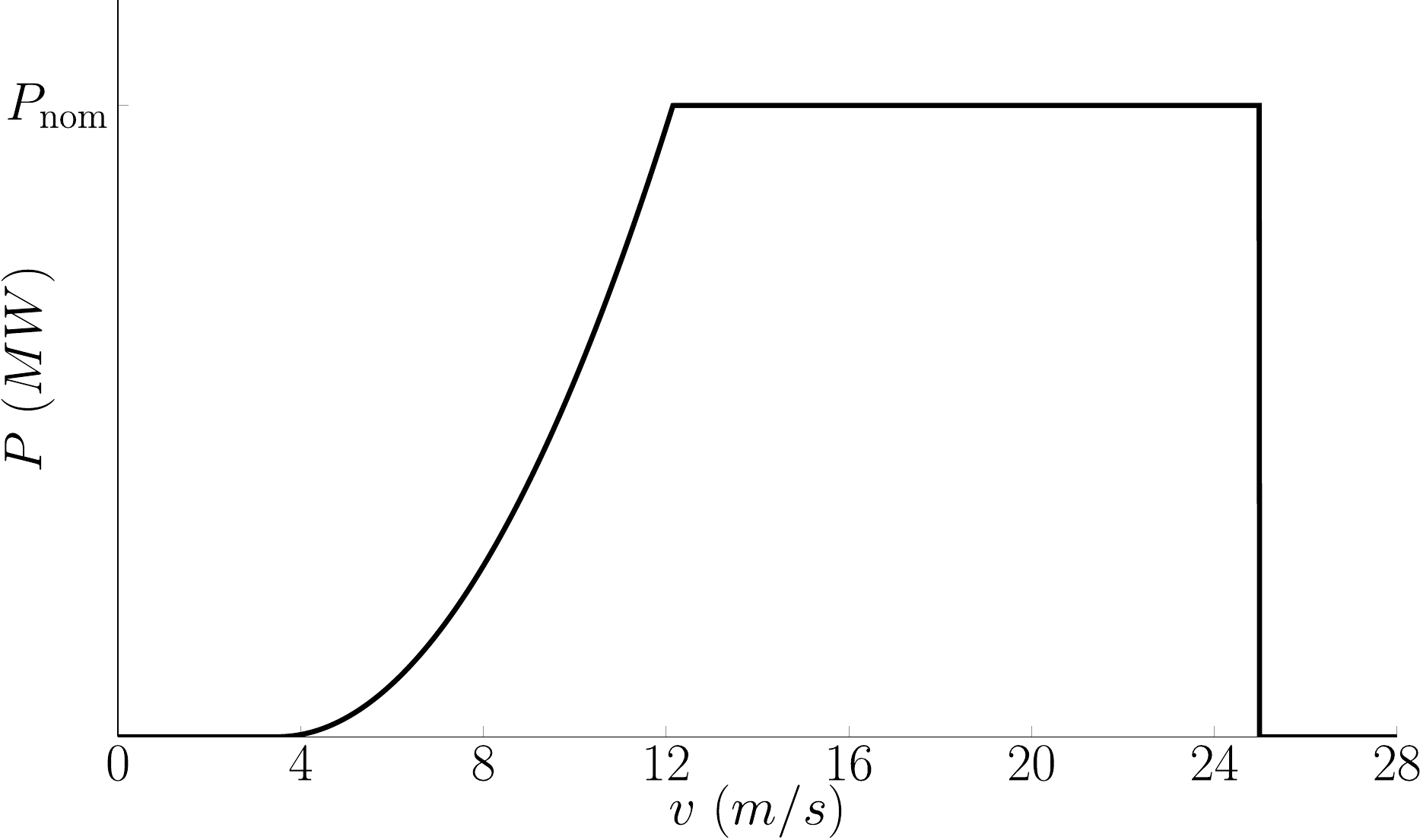}
\caption{Power curve of a wind generator.}
\label{fig:p_curve}
\end{figure}

\paragraph{Speed and power level of wind generators}
The uncertainty about the production level of wind turbines is inherited from the uncertainty about the wind speed. The stochastic process that we consider governs the wind speed, which is assumed to be uniform across the network. The production level of the wind generators is then obtained by using a deterministic function that depends on the wind speed realization, this function is the power curve of the considered generator. We can formulate this phenomenon as:
\begin{align}
v_{t+1} &= f_v(v_{t}, \dots, v_{t-N_{v}+1}, \bm{s}^{(aux)}_t, \bm{w}^{(v)}_t)\,,\label{eq:trans_ws}\\
P_{g,t+1} &= \eta_g(v_{t+1}), \forall g \in \text{wind generators} \subset \mathcal{G}\,,
\end{align}
such that $\bm{w}^{(v)}_{t} \sim p_{\mathcal{W}^{(v)}}(\cdot)$ denotes some components of $\bm{w}_t \sim p_{\mathcal{W}}(\cdot)$ and where $\eta_g$ is the power curve of generator $g$. A typical example of power curve $\eta_g(v)$ is illustrated in Fig.~\ref{fig:p_curve}.

\paragraph{Irradiance and photovoltaic production}

Like wind generators, the photovoltaic generators inherit their uncertainty in production level from the uncertainty associated with their energy source. This source is represented by the level of solar irradiance, which is the power level of the incident solar energy per square meter. The irradiance level is the stochastic process that we model, while the production level is obtained by a deterministic function of the irradiance and of the surface of photovoltaic panels. This function is simpler than the power curve of wind generators and is defined as
\[
P_{g,t} = \eta_g \cdot \text{surf}_g \cdot ir_t\,,
\]
where $\eta_g$ is the efficiency factor of the panels, assumed constant and with a typical value around $15\%$, while $\text{surf}_g$ is the surface of the panels in $m^2$ and is specific to each photovoltaic generator. The irradiance level is denoted by $ir_t$ and the whole phenomenon is modeled by the following process:
\begin{align}
ir_{t+1} &= f_{ir}(ir_{t}, \dots, ir_{t-N_{ir}+1}, \bm{s}^{(aux)}_t, \bm{w}^{(ir)}_t)\,,\label{eq:trans_ir}\\
P_{g,t+1} &= \eta_g \cdot \text{surf}_g \cdot ir_{t+1}, \forall g \in \text{solar generators} \subset \mathcal{G}\,,
\end{align}
such that $\bm{w}^{(ir)}_{t} \sim p_{\mathcal{W}^{(ir)}}(\cdot)$ denotes some components of $\bm{w}_t \sim p_{\mathcal{W}}(\cdot)$.

\paragraph{Auxiliary modeling variables}

The evolution of auxiliary modeling variables depends on their meaning and must be determined when instantiating the presented abstract decision model. The evolution of each component of $\bm{s}^{\mathrm{(aux)}}_t$ can be either stochastic or deterministic and, without loss of generality, we can write:
\begin{align}
\bm{s}^{\mathrm{(aux)}}_{t+1} = f_{aux}(\bm{s}_t, \bm{w}^{(\mathrm{aux})}_t)\,,
\end{align}
such that $\bm{w}^{(\mathrm{aux})}_t \sim p_{\mathcal{W}^{(\mathrm{aux})}}(\cdot)$ denotes some components of $\bm{w}_t \sim p_{\mathcal{W}}(\cdot)$.

\paragraph{Impact of control actions}
The stochastic processes that we described govern the evolution of the state $\bm{s}^{(1)}_t \in \mathcal{S}^{(1)}$ of the consumption of loads (flexibility services excluded) and of the power level of energy sources of DGs. We now define the evolution of the components of $\bm{s}^{(2)}_t \in \mathcal{S}^{(2)}$ by integrating the control actions of the DSO. Concerning the modulation instructions of the generators, we have:
\begin{align}
\forall g \in \mathcal{G} : &&\overline{P}_{g,t+1} &=
	\begin{cases}
		\overline{p}_{g,t}	& \text{if } (\overline{p}_{g,t},\hat{q}_{g,t}) \in \mathcal{O}_g\,,\\
		\max_{(p,\hat{q}_{g,t}) \in \mathcal{O}_g} p					& \text{otherwise}\,,
\end{cases}\label{eq:p_max}\\
\forall g \in \mathcal{G} : &&\hat{Q}_{g,t+1} &= \hat{q}_{g,t}\,,
\end{align}
where $\max_{(p,\hat{q}_{g,t}) \in \mathcal{O}_g} p$ denotes the maximal active production level that generator $g$ can sustain with a reactive set-point of $\hat{q}_{g,t}$. It is used if needed to ensure that the instructed $(\overline{P}_{g,t+1},\hat{Q}_{g,t+1})$ point is a valid operating point, such a defined by equation~(\ref{eq:valid_pq}). As for the components dedicated to the flexible loads, their evolution is governed by:
\begin{align}
\forall d \in \mathcal{F} : &&flex_{d,t+1} &= \max(flex_{d,t}-1~;~0) + act_{d,t} T_d\,,\label{eq:flex_trans}\\
\forall d \in \mathcal{F} : &&\Delta	 P_{d,t+1} &=
	\begin{cases}
		\Delta P_d(T_d - flex_{d,t+1} + 1)	& \text{if } flex_{d,t+1} > 0\\
		0     	& \text{if } flex_{d,t+1} = 0\,.
	\end{cases}\label{eq:mod_sig}
\end{align}
From vectors $\bm{s}^{(1)}_t$ and $\bm{s}^{(2)}_t$, we can determine the active and reactive power injections at nodes and thus obtain the value of the electrical quantities at nodes $n \in \mathcal{N}$ and links $(m,n) \in \mathcal{L}$ of the network:
\begin{align}
P_{n,t} &= \sum_{g \in \mathcal{G}(n)} \hspace{-0.8em} \min(\overline{P}_{g,t}; P_{g,t}) + \sum_{d \in \mathcal{C}(n)} P_{d,t} + \hspace{-0.8em} \sum_{d \in \mathcal{F}(n)} \hspace{-0.3em} \Delta P_{d,t}\,,\label{eq:pf0}\\
Q_{n,t} &=  \sum_{g \in \mathcal{G}(n)} \hspace{-0.8em} \hat{Q}_{g,t} + \hspace{-0.5em} \sum_{d \in \mathcal{C}(n)} \hspace{-0.5em} Q_{d,t} + \hspace{-0.8em} \sum_{d \in \mathcal{F}(n)} \hspace{-0.3em} \tan{\phi_d} \Delta P_{d,t}\,,\label{eq:pf1}\\
P_{n,t} - jQ_{n,t} &= V^*_{n,t} \bm{Y}_{n\cdot} V_{n,t}\,,\label{eq:pf2}\\
|I_{mn,t}| &= \left|\left(|t_{mn}|^2 V_{m,t} -(t^{(l)}_{mn})^* t^{(l)}_{nm} V_{n,t}\right)Y^{(br)}_{mn}\right|\,.\label{eq:last_f_def}
\end{align}

\subsubsection{Reward function and goal}
\label{sec:reward}

In order to evaluate the performance of a policy, we first specify the reward function $r : \mathcal{S} \times \mathcal{A}_{\bm{s}} \times \mathcal{S} \to \mathbb{R}$, which associates an instantaneous reward for each transition of the system from a period $t$ to a period $t+1$:
\begin{align}
r(\bm{s}_{t}, \bm{a}_t, \bm{s}_{t+1}) = &- \underbrace{\sum_{g\in\mathcal{G}} \max\{0,\frac{P_{g,t+1}-\overline{P}_{g,t+1}}{4}\} C_g^{curt}(\bm{s}_{t+1}^{(\mathrm{aux})})}_{\text{curtailment cost of DGs}} \notag \\ &- \underbrace{\sum_{d\in \mathcal{F}} act_{d,t} C^{flex}_d}_{\substack{\text{activation cost}\\\text{of flexible loads}}} - \underbrace{\Phi(\bm{s}_{t+1})}_{\text{penalty function}}\,,\label{eq:reward_fct}
\end{align}
where $C_g^{curt}(\cdot)$ is a per-generator function that defines the curtailment price, while $C_d^{flex}(\cdot)$ defines the activation cost for each flexible load. In this generic definition of the reward, we allow both functions to depend on the auxiliary state variables so that it can model arbitrary processes.
The function $\Phi$ aims at penalizing a policy that leads the system into an undesirable state (e.g. that violates the operational limits or induces many losses) and, together with $C_g^{curt}$ and $C_d^{flex}$, it must be defined when instantiating the decision model. Note that equation~(\ref{eq:reward_fct}) is such that the higher the operational costs and the larger the violations of operational limits, the more negative the reward function.

We can now define the \emph{return over T periods}, denoted $R_T$, as the weighted sum of the rewards that are observed over a system trajectory of $T$ periods
\begin{align}
R_T = \sum_{t=0}^{T-1} \gamma^{t} r(\bm{s}_{t}, \bm{a}_t, \bm{s}_{t+1})\,,
\end{align}
where $\gamma \in ]0;1[$ is the discount factor. Given that $\gamma^t < 1$ for $t > 0$, the further in time the transition from period $t=0$, the less importance is given to the associated reward. Because the operation of a DN must always be ensured, it does not seem relevant to consider returns over a finite number of periods and we introduce the \emph{return} $R$ as
\begin{align}
R = R_{\infty} = \lim\limits_{T \to \infty} \sum_{t=0}^{T-1} \gamma^{t} r(\bm{s}_{t}, \bm{a}_t, \bm{s}_{t+1})\,,
\end{align}
that corresponds to the weighted sum of the rewards observed over an infinite trajectory of the system. Given that the costs have finite values, assuming the same for penalties, and observing that the reward function $r$ is the sum of an infinite number of these costs and penalties, a constant $C$ exists such that, $\forall (\bm{s}_{t}, \bm{a}_t, \bm{s}_{t+1})\in\mathcal{S}\times\mathcal{A}_{\bm{s}}\times\mathcal{S}$, we have $|r(\bm{s}_{t}, \bm{a}_t, \bm{s}_{t+1})| < C$ and thus
\begin{align}
|R| < \lim\limits_{T \to \infty} C\sum_{t=0}^{T-1} \gamma^{t} = \frac{C}{1-\gamma}\,.
\end{align}
It means that even if the return R is defined as an infinite sum, it converges to a finite value. One can also observe that, because $\bm{s}_{t+1} = f(\bm{s}_t, \bm{a}_t, \bm{w}_t)$, a function $\rho : \mathcal{S} \times \mathcal{A} \times \mathcal{W} \to \mathbb{R}$ exists that aggregates functions $f$ and $r$ such that
\begin{align}
\rho(\bm{s}_t, \bm{a}_t, \bm{w}_t) = r(\bm{s}_t, \bm{a}_t, f(\bm{s}_t, \bm{a}_t, \bm{w}_t)) = r(\bm{s}_t, \bm{a}_t, \bm{s}_{t+1})\,,
\end{align}
with $\bm{w}_t \sim p_{\mathcal{W}}(\cdot)$. Let $\pi : \mathcal{S} \to \mathcal{A}_{\bm{s}}$ be a policy that associates a control action to each state of the system. We can define, starting from an initial state $\bm{s}_0 = \bm{s}$, the expected return $R$ of the policy $\pi$ by
\begin{align}
J^{\pi}(\bm{s}) = \lim\limits_{T \to \infty}\hspace{0.75em}\underset{\underset{\scriptscriptstyle t=0,1,\dots }{\scriptscriptstyle \bm{w}_t \sim p_{\mathcal{W}}(\cdot)}}{\mathbb{E}} \{\sum_{t=0}^{T-1} \gamma^{t} \rho(\bm{s}_t, \pi(\bm{s}_t), \bm{w}_t) | \bm{s}_0 = \bm{s}\}\,.
\end{align}
We denote by $\Pi$ the space of all the policies $\pi$. For a DSO, addressing the operational planning problem described in Section~\ref{sec:probStatement} is equivalent to determine an optimal policy $\pi^*$ among all the elements of $\Pi$, i.e. a policy that satisfies the following condition
\begin{align}
J^{\pi^*}(\bm{s}) \geq J^{\pi}(\bm{s}), \forall \bm{s} \in \mathcal{S}, \forall \pi \in \Pi\,.
\end{align}
It is well known that such a policy satisfies the Bellman equation \cite{bellman1957dp}, which can be written
\begin{align}
J^{\pi^*}(\bm{s}) = \max_{\bm{a} \in \mathcal{A}_{\bm{s}}} \underset{\scriptscriptstyle \bm{w}\sim p_{\mathcal{W}}(\cdot)}{\mathbb{E}}\big\{ \rho(\bm{s}, \bm{a}, \bm{w}) + \gamma J^{\pi^*}(f(\bm{s},\bm{a},\bm{w})) \big\}, \forall \bm{s} \in \mathcal{S}\,.\label{eq:bell}
\end{align}
If we only take into account the space of stationary policies (i.e. that selects an action independently of time $t$), it is without loss of generality comparing to the space of policies  $\Pi' : \mathcal{S} \times \mathcal{T} \to \mathcal{A}$ because the return to be maximized corresponds to an infinite trajectory of the system \cite{bertsekas1978stochastic}.

\section{Lookahead optimization model}
\label{sec:solution}

We now describe a look-ahead algorithm to build a policy based on stochastic programming.
The principle is, at each time step $t \in \mathcal{T}$, to optimize a model $\mathcal{M}_t$ of the system over a finite time horizon $\mathcal{T}_t = \{t, ..., t+T-1\}$ and to apply the control action $\hat{\bm{a}}^*_t = \hat{\pi}^*_{\mathcal{M}_t}(\bm{s}_t)$ that corresponds to the first stage of the model. This approximate optimal policy $\hat{\pi}^*_{\mathcal{M}_t}$  can be formulated as
\begin{eqnarray}
\hat{\pi}^*_{\mathcal{M}_t}(\bm{s}_t) =
\underset{\bm{a}_t}{\arg} \underset{
	\begin{array}{c}
	\scriptstyle \bm{s}_{\scriptscriptstyle t'}\,,\scriptstyle \bm{a}_{\scriptscriptstyle t'}\\[-0.4em]
	\scriptscriptstyle \forall t' \in \mathcal{T}_t
	\end{array}
}{\max}& &
\underset{\scriptstyle \bm{w}_{t'} {\scriptscriptstyle\sim} p_{\scriptscriptstyle\mathcal{W}}(\cdot)}{\mathbb{E}}\Big[\sum_{t'=t}^{t+T-1} \gamma^{t'-t} r(\bm{s}_{t'}, \bm{a}_{t'}, f(\bm{s}_{t'}, \bm{a}_{t'}, \bm{w}_{t'}))\Big]\label{eq:mp1}\\
\text{s.t.}\hspace{0.5em}	& & \bm{s}_{t'} = f(\bm{s}_{t'-1}, \bm{a}_{t'-1}, \bm{w}_{t'-1})\,,\quad \forall t' \in \mathcal{T}_t \backslash \{t\}\\
& & \bm{a}_{t'} \in \mathcal{A}_{\bm{s}_{t'}}\,,\quad \forall t' \in \mathcal{T}_t \,,\label{eq:mp2}
\end{eqnarray}
where the shorter the horizon $T$, the higher the approximation error.

The finite lookahead time horizon is not the only source of approximation. First, there is no exact numerical method to solve (\ref{eq:mp1})-(\ref{eq:mp2}) without requiring a discrete approximation of the continuous stochastic processes \cite{Shapiro2009}. We detail in Section~\ref{sec:scen_tree} how to build such a discrete approximation. Then, because of the nonlinearity of power-flow equations on the one hand, and the integer variables that model the activation of flexibility services on the other hand, the resulting mathematical problem is very complex to solve. For this reason, it is often required either to resort to local optimization techniques and heuristics, or to use relaxations and approximations of the power-flow equations. In particular, we describe in Section~\ref{sec:net_models} several models of the electrical network of different complexity and accuracy.

\subsection{Model instantiation}
\label{sec:mdp_inst}

The decision model presented in Section~\ref{sec:dynamics} is generic on some of its elements. We now instantiate these elements to obtain a practical model that can be implemented to perform numerical simulations.

\paragraph{Auxiliary state variable}

We limit the vector $\bm{s}_t^{(\mathrm{aux})}$ to a single auxiliary variable that indicates the time of the day:
\begin{align}
	\bm{s}_t^{(\mathrm{aux})} = q_{t}\,,
\end{align}
which takes values in $\{0,\dots,95\}$ to identify the quarter of an hour in the day. This information will be used as an input of the modulation price functions and of the transition function of both production and consumption processes. The relation that governs the evolution of $q_t$ can be stated as a function $f_{aux}: \{0,\dots,95\} \mapsto \{0,\dots,95\}$, which is defined as:
\begin{align}
q_{t+1} = f_{aux}(q_{t}) = (q_t + 1) \bmod 96 \quad.
\end{align}

\paragraph{Modulation prices}

For the sake of simplicity, we consider that the curtailment price functions $C_g^{curt}$ depend exclusively on $q_t$. The time of the day being deterministic, these functions are deterministic too and correspond to arrays of 96 price values, which span a whole day. Concerning the activation costs $C^{flex}_d$, they are assumed to be constant on a per-load basis. The values of both the arrays and constants are specified in Section~\ref{sec:test} when presenting the test instances.

\paragraph{Penalty function}

We choose to penalize a policy for violating operational limits and for the active losses in the network. This is implemented using the following function:
\begin{align}
\Phi(\bm{s}_{t+1}) =~&k.\big(\sum_{n \in \mathcal{N}} [\max(0,|V_{n,t+1}| - \overline{V}_{n}) + \max(0,\underline{V}_{n} - |V_{n,t+1}|)] \nonumber\\
&\hspace{1em}+ \sum_{(m,n) \in \mathcal{L}} \max(0,|I_{mn,t+1}| - \overline{I}_{mn})\,\big) \nonumber\\
&+ C_{loss}(q_{t+1}) \sum_{n\in\mathcal{N}} \frac{P_{n,t+1}}{4}
\,,\label{eq:barrier}
\end{align}
where $|V_{n,t+1}|$ ($n \in \mathcal{N}$) and $I_{mn,t+1}$ ($(m,n) \in \mathcal{L}$) are determined from $\bm{s}_{t+1}$ using equations~(\ref{eq:pf0})-(\ref{eq:last_f_def}), and where $k \in \mathbb{R}_0^+$ is a typically large constant. The per-unit price $C_{loss}(q_{t+1})$ of losses is a deterministic function of the quarter of hour and corresponds to an array of 96 price values.

\paragraph{Production and consumption processes}

The instantiated versions of transition functions (\ref{eq:trans_cons}), (\ref{eq:trans_ws}), and (\ref{eq:trans_ir}), of the stochastic quantities (i.e. the consumption of the loads, the wind speed, and the level of solar irradiance) have the following structure:
\begin{align}
	x_{t+1} &= f_x(x_{t},\dots,x_{t-N_x+1}, q_t, w_t^{(x)})\,,\\
	&= \mu_{x,t+1} + \sigma_{x,t+1} \cdot w_t^{(x)}\,,\\
	\text{with }w_t^{(x)} &\sim p_{\mathcal{W}^{(x)}}\left(\cdot|\frac{x_{t}-\mu_{x,t}}{\sigma_{x,t}},\dots,\frac{x_{t-N_x}-\mu_{x,t-N_x+1}}{\sigma_{x,t-N_x+1}}\right)\,,\label{eq:cond_prob} 
\end{align}
where $x$ denotes the considered process, and where $\mu_{x,t \pm \Delta t}$ and $\sigma_{x,t \pm  \Delta t}$ are shortcuts for the following per-process functions:
\begin{align}
	\mu_{x,t \pm \Delta t} &= \mu_{x}\Big((q_t \pm \Delta t) \bmod 96\Big)\,,\\
	\sigma_{x,t \pm \Delta t} &= \sigma_{x}\Big((q_t \pm \Delta t) \bmod 96\Big)\,.
\end{align}
These functions normalize the processes and remove their diurnal seasonality, and the conditional distribution of $w_t^{(x)}$ is then assumed to be stationary. The details of the conditional density functions are not required for the development of the lookahead optimization model, we specify in Section~\ref{sec:test} a possible procedure to learn these functions from time series of measurements.

\subsection{Discretization of the random process}
\label{sec:scen_tree}

\begin{figure}[hb] 
	\centering
	\includegraphics[width=3.25in]{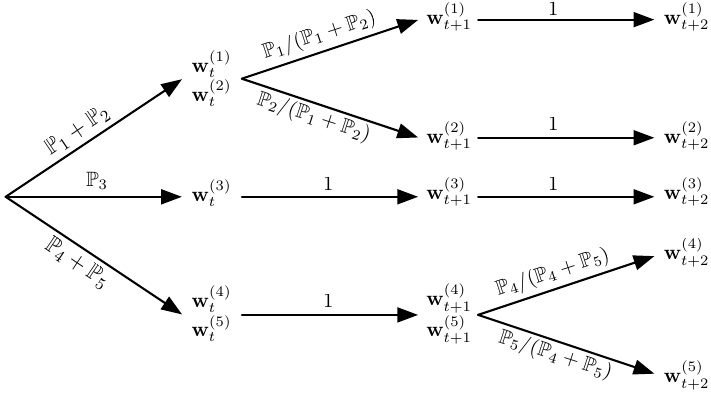}
	\caption{Example of scenario tree with $T = 3$ and $W = 5$.}
	\label{fig:ex_scen_tree}
\end{figure}

The random process needs to be discretized over the look-ahead horizon to implement the policy with a computer program. A prevalent technique is to use a scenario tree \cite{defourny2011multistage} for this purpose.  At each time step $t \in \mathcal{T}$, the evolution of the stochastic components is aggregated as a finite set $\tilde{\mathcal{W}}^T_t$ of outcome trajectories of the exogenous variables:
\begin{align}
\tilde{\mathcal{W}}^T_t = \{(\bm{w}^{(k)}_{t}, \dots, \bm{w}^{(k)}_{t+T-1}) | k =1,\dots,W\}\,,
\end{align}
and a probability $\mathbb{P}_k$ is associated to each trajectory $k \in \{1,\dots,W\}$. If two trajectories $i$ and $j$ share the same outcomes up to stage $o$, i.e. if $(\bm{w}^{(i)}_{t'},\dots,\bm{w}^{(i)}_{t'+o}) = (\bm{w}^{(j)}_{t'},\dots,\bm{w}^{(j)}_{t'+o})$, they can be interpreted as a single trajectory of probability $\mathbb{P}_i + \mathbb{P}_j$ up this stage. Fig.~\ref{fig:ex_scen_tree} provides an example of such a scenario tree, where the nodes represent the outcomes and the edges correspond to the transition probabilities.

\subsection{Mathematical program}

The purpose of the first term in the penalty function is to be an incentive to prevent the policy to bring the system in a state that violates operational limits. This definition allows to evaluate any kind of policy. In a mathematical programming setting, we remove this term from the objective function and add  operational constraints defined in equations~(\ref{eq:v_limits}) and (\ref{eq:i_max}) to (\ref{eq:mp1})-(\ref{eq:mp2}). The new objective function becomes:
\begin{align}
\mathrm{cost}(\bm{s}_t, \bm{a}_t, \bm{s}_{t+1}) = &\sum_{g\in\mathcal{G}} \max\{0,\frac{P_{g,t+1}-\overline{P}_{g,t+1}}{4}\} C^{curt}_g(q_{t+1}) + \sum_{d\in \mathcal{F}} act_{d,t} C^{flex}_d\nonumber\\
&+ C_{loss}(q_{t+1}) \sum_{n\in\mathcal{N}} \frac{P_{n,t+1}}{4}\,.\label{eq:cost_fct}
\end{align}

Taking into account the discretization of the stochastic processes, the objective function defined in equation~(\ref{eq:cost_fct}), and the additional constraints, we can formulate a new approximate optimal policy $\hat{\pi}^*_{\hat{\mathcal{M}}_t}$ as
\begin{eqnarray}
\hat{\pi}^*_{\hat{\mathcal{M}}_t}(\bm{s}_t) =
\underset{\bm{a}_t}{\arg} \hspace{-1em} \min_{
\begin{array}{c}
\scriptstyle \bm{s}_{\scriptscriptstyle t}^{\scriptscriptstyle(k)},\dots,\bm{s}_{\scriptscriptstyle t+T}^{\scriptscriptstyle(k)},\\[-0.1em]
\scriptstyle \bm{a}_{\scriptscriptstyle t}^{\scriptscriptstyle(k)},\dots,\scriptstyle \bm{a}_{\scriptscriptstyle t+T-1}^{\scriptscriptstyle(k)},\\[-0.3em]
\scriptscriptstyle \forall k \in \{1,\dots,\mathcal{W}\}
\end{array}}\hspace{-2em} & &
\hspace{1em} \sum_{k=1}^{W} \sum_{t'=t}^{t+T-1} \Big[\mathbb{P}_k \gamma^{t'-t} \mathrm{cost}(\bm{s}^{(k)}_{t'}, \bm{a}^{(k)}_{t'}, \bm{s}^{(k)}_{t'+1})\Big] \label{eq:msp1} \\
\text{s.t.}	& & \forall t' \in \mathcal{T}_t: \notag\\
		& & \quad \bm{a}^{(i)}_{t'} = \bm{a}^{(j)}_{t'}, \quad \forall i,j~\text{s.t.}~{\scriptstyle(\bm{w}^{(i)}_{t},\dots,\bm{w}^{(i)}_{t'}) = (\bm{w}^{(j)}_{t},\dots,\bm{w}^{(j)}_{t'})}\,,\label{eq:nonanticip}\\
		& & \forall (k,t') \in \{1,\dots,W\} \times \mathcal{T}_t : \notag\\
		& & \quad \bm{s}^{(k)}_{t'+1} = f(\bm{s}^{(k)}_{t'}, \bm{a}^{(k)}_{t'}, \bm{w}^{(k)}_{t'})\,,\label{eq:minlp_s}\\
		& & \quad \bm{a}^{(k)}_{t'} \in \mathcal{A}_{\bm{s}^{(k)}_{t'}}\,,\label{eq:minlp_a}\\
		& & \forall (n,k,t') \in \mathcal{N} \times \{1,\dots,W\} \times \mathcal{T}_t : \notag\\
		& & \quad \underline{V}^{(k)}_{n,t'} \leq |V^{(k)}_{n,t'}| \leq \overline{V}^{(k)}_{n,t'}\,,\\
		& & \forall (m,n,k,t') \in \mathcal{L} \times \{1,\dots,W\} \times \mathcal{T}_t : \notag\\
		& & \quad |I_{mn,t}| \leq \overline{I}_{mn}\,,\label{eq:msp2}
\end{eqnarray}
where (\ref{eq:minlp_s}) stands for equations (\ref{eq:trans_cons})-(\ref{eq:last_f_def}), and (\ref{eq:minlp_a}) for equation (\ref{eq:feas_actions}). The model takes into account that decisions at stage $t' \in \mathcal{T}_t$ only depend on exogenous information up to stage $t'$, i.e. that future unknown data is not used, which is why we integrate the nonanticipativity constraints \cite{Shapiro2009} to the mathematical program using equation (\ref{eq:nonanticip}). Problem (\ref{eq:msp1})-(\ref{eq:msp2}) is a  mixed-integer program (MIP). For a given distribution system, its complexity depends mainly on the network model chosen to represent power-flow equations (cf. Section~\ref{sec:net_models}), and on the number of scenarios representing the uncertainty.

\subsection{Detailed model of control actions}

Implementing the model of control actions of Section~\ref{sec:control_actions} in a mathematical program is not straightforward. We now present how it can be implemented in problem (\ref{eq:msp1})-(\ref{eq:msp2}).

\subsubsection{Generation curtailment}
\label{sec:MIP_model_curtailment}

This section focuses on the curtailment decision model. To ease the reading, we focus on one particular generator and thus omit subscript $g$. Note first that the active power injection term of equation~(\ref{eq:pf0}) that follows a curtailment instruction $\overline{p}_{t} \geq 0$ from stage $t$, i.e. $\min(\overline{P}_{t+1}; P_{t+1})$, is translated in the mathematical program by
\begin{align}
\min(\overline{P}_{t+1}; P_{t+1}) = P_{t+1} - p_{curt,t+1}\,,
\end{align}
where $p_{curt,t+1} \geq 0$ is the amount of active power curtailment induced by the power limit instruction. This quantity is easy to determine in a deterministic setting:
\begin{align}
	\overline{P}_{t+1} &=  \overline{p}_{t}\,,\\
	p_{curt,t+1} &= P_{t+1} - \overline{P}_{t+1}\,,
\end{align}
where $P_{t+1}$ is the potential active production level at time step $t+1$.

Considering several scenarios leads to a less obvious definition of the amount of curtailment. Let $\overline{p}_{t}$ denote the curtailment instruction that, at time $t$, is shared by all scenarios $j \in \{1,\dots,W\}$ such that $(\dots,\bm{w}^{(k)}_{t-1},\bm{w}^{(k)}_{t}) = (\dots,\bm{w}^{(j)}_{t-1},\bm{w}^{(j)}_{t})$. For one generator, the maximum power allowed and the curtailed power are defined by the following set of constraints for each scenario $k$:
\begin{align}
\overline{P}^{(k)}_{t+1} &= \overline{p}_{t}\,,\\
\Delta p^{(k)}_{t+1} &= P_{t+1}^{(k)} -  \overline{P}^{(k)}_{t+1}\,,\\
p^{(k)}_{curt,t+1} &= \max(0, \Delta p^{(k)}_{t+1} \label{eq:p_curt_i})\,,
\end{align}
where $P^{(k)}_{t+1}$ is the potential active production level in scenario $k$, and $\Delta p^{(k)}_{t+1}$ is an auxiliary variable that has no physical meaning, since it can be negative when $\overline{p}_{t}$ is not restrictive for scenario $k$ (i.e. when $\overline{p}_{t} \geq P^{(k)}_{t+1}$). The variable $p_{curt,t+1}^{(k)}$ would be the power curtailed if scenario $k$ realizes. It contributes linearly to the value of the objective function, proportional to the curtailment cost and weighted by the probability of scenario $k$. A common relaxation of the $\max$ operator of constraint \eqref{eq:p_curt_i} for a variable that tends to be minimized is
\begin{align}
p^{(k)}_{curt,t+1} &\geq \Delta p_{t+1}^{(k)}\,. \label{eq:UB_pcurt}
\end{align}
However, this holds only if constraint \eqref{eq:UB_pcurt} is always tight, which is not always true in problem (\ref{eq:msp1})-(\ref{eq:msp2}). Without preventing $p^{(k)}_{curt,t+1}$ to be greater than $\Delta p^{(k)}_{t+1}$, it would allow to discriminate the curtailment decisions between different scenarios even though the nonanticipativity constraint (\ref{eq:nonanticip}) is respected. Indeed, the amount of power curtailed could be increased beyond $P_{t+1}^{(k)} - \overline{P}^{(k)}_{t+1}$ and it would differ from the set point $\overline{p}_{t}$, which is guaranteeing nonanticipativity by being common for all subsequent scenarios.

From this analysis, we conclude that a continuous implementation of equation~(\ref{eq:p_curt_i}) is not possible and we model it using equation~(\ref{eq:UB_pcurt}) and the following additional constraints:
\begin{align}  
p^{(k)}_{curt,t+1} &\leq \Delta p^{(k)}_{t+1} - y^{(k)} \underline{\Delta P}^{(k)}_{t+1} \label{eq:LB_pcurt_1}\,,\\
p^{(k)}_{curt,t+1} &\leq  (1 - y^{(k)}_{t+1}) \ \overline{\Delta P}^{(k)}_{t+1} \label{eq:LB_pcurt_2}\,,\\
y^{(k)}_{t+1} &\in \{0,1\}\,,
\end{align}
where $\overline{\Delta P}^{(k)}_{t+1}$ and $\underline{\Delta P}^{(k)}_{t+1}$ are parameters that indicate the maximal and minimal values that $\Delta p_{t+1}^{(k)}$ can take, respectively. It corresponds to a \textit{big M} formulation \cite{fortuny1981representation} and can be interpreted as follow:
\begin{itemize}
	\item  if $\Delta p^{(k)}_{t+1} < 0$, constraint \eqref{eq:LB_pcurt_1} is satisfied only if $y^{(k)}_{t+1} = 1$, and constraint \eqref{eq:LB_pcurt_2} then forces $p_{curt,t+1}^{(k)} = 0$ ;
	\item if $\Delta p^{(k)}_{t+1} \geq 0$, constraints \eqref{eq:UB_pcurt} and \eqref{eq:LB_pcurt_1} can be satisfied simultaneously only if $y^{(k)}_{t+1} = 0$ .
\end{itemize}
Parameters $\underline{\Delta P}^{(k)}_{t+1}$ and $\overline{\Delta P}^{(k)}_{t+1}$ should be chosen such that the continuous relaxation is as tight as possible. For instance for a wind turbine $\overline{\Delta P}^{(k)}_{t+1} = P^{(k)}_{t+1}$ and $\underline{\Delta P}^{(k)}_{t+1} = P^{(k)}_{t+1} - \max_j P^{(j)}_{t+1}$.

Finally, the curtailment instruction $\overline{p}_{g,t}$ are recovered from the solution of $\hat{\mathcal{M}}_t$ upon the following processing of the solution:
\begin{align}
\forall g \in \mathcal{G}: \overline{p}_{g,t} \leftarrow \begin{cases}
\overline{p}_{g,t} & \text{if}~\exists k \in \{1,...,W\}~\text{s.t.}~\overline{p}_{g,t} < P^{(k)}_{g,t+1}\,,\\
+ \infty & \text{otherwise}\,,
\end{cases} \label{eq:proc_curt}
\end{align}
This processing is introduced because, in $\hat{\mathcal{M}}_t$, the value of $\overline{p}_{g,t}$ has no meaning when it does not induce an actual curtailment for at least one scenario. Therefore, it makes no sense to interpret these variables as curtailment instructions and equation~(\ref{eq:proc_curt}) makes sure that curtailment actions sent to the system actually corresponds to curtailment decisions in the optimization model.

\subsubsection{Activation of flexibility services}
This section details how the control actions $act_{d,t}$ defined in Section~\ref{sec:control_actions} are computed. To ease the reading, we focus on one particular device and thus omit subscript $d$ in this section. The superscript $(k)$ is also dropped and the following equations simply needs to be repeated for each scenario of the lookahead model, with equation~(\ref{eq:nonanticip}) guaranteeing the nonanticipativity of the model.  We first define several auxiliary variables:
\begin{itemize}
	\item $z_t \in \{0,1\} $ is a binary variable used to model the $\max$ operator for state transitions;
	\item $m_t \in \mathbb{Z}_+$ is a integer variable used for state transitions.
\end{itemize}
Ignoring the activation signal, the transition rule of the flexible state, i.e. $$m_{t+1} = \max(0,flex_{t}-1)\,,$$ is implemented through the following \textit{big M} formulation:
\begin{align}
m_{t+1} &\leq flex_{t}-1 + z_t \label{eq:flex_max_1}\,\\
m_{t+1} &\leq T \left(1-z_t\right) \label{eq:flex_max_2}\,\\
m_{t+1} &\geq flex_{t} - 1 \label{eq:flex_max_3}\,.
\end{align}
The influence of the activation signal is then incorporated to the flexible state:
\begin{align}
flex_{t+1} = m_{t+1} + act_t T\,,
\end{align}
while the following constraint prevents a double activation of a flexibility service:
\begin{align}
a_t + \frac{flex_t}{T} &\leq 1\,.
\end{align}
Finally, the value of the effective modulation signal, defined in equation~(\ref{eq:mod_sig}), is implemented as:
\begin{align}
\Delta P_t = \sum_{t' : t-t' \leq T} a_{t-t'} \Delta P(t-t') \label{eq:flex_delta}\,,
\end{align}
where $\Delta P(\cdot)$ is the modulation curve of the load, which produces parameters for the mathematical program.

\subsection{Detailed network models}
\label{sec:net_models}

We detail how the network model described in Section~\ref{EDSmodel} is precisely instantiated in the AC non-convex case, then we describe a linearization approach and finally a second order cone program (SOCP) model. To ease reading we consider only one time step and omit the subscripts $t$ and the scenario notation, but in reality these equations are replicated for each time step or node of the scenario tree.
In this section, we define $$g_{mn}+jb_{mn} = Y^{(br)}_{mn}$$ and $$g^{(sh)}_m+jb_m^{(sh)} = \sum_{n:(m,n)\in \mathcal{L}}Y^{(sh)}_{mn}.$$
We also consider arbitrarily that node $1$ is a slack bus which sets a reference phase angle of $0$ and a fixed voltage magnitude.

\subsubsection{Non-convex AC model}
\label{sec:AC-model}
We chose to express relations \eqref{eq:pf_eqs} in rectangular coordinates. Hence we define variables 
\begin{itemize}
	\item $e_n$ as the real part of $V_n$
	\item $f_n$ as the imaginary part of $V_n$
	\item $P_{mn}$ as the active power leaving bus $m$ and flowing in link $(m,n)$
	\item $Q_{mn}$ as the reactive power leaving bus $m$ and flowing in link $(m,n)$
	\item $P^{shunt}_n$ as the active power shunted at bus $n$
	\item $Q^{shunt}_n$ as the reactive power shunted at bus $n$.
\end{itemize}
The above powers are defined as
\begin{align}
\forall (m,n) \in &\mathcal{L}: \notag \\
P_{mn} &= e_m \left(g_{mn} (e_m-e_n) - b_{mn} (f_m-f_n)\right) + f_m \left(b_{mn} (e_m-e_n) + g_{mn} (f_m-f_n)\right), \label{eq:Pmn_def_AC}\\
Q_{mn} &= f_m \left(g_{mn} (e_m-e_n) - b_{mn} (f_m-f_n)\right) - e_m \left(b_{mn} (e_m-e_n) + g_{mn} (f_m-f_n)\right), \label{eq:Qmn_def_AC} \\ 
\forall n \in &\mathcal{N}: \notag \\
P^{shunt}_n &= g_n^{(sh)} (e_n^2 + f_n^2) \label{eq:Pshunt_def_AC}, \\ 
Q^{shunt}_n &= b_n^{(sh)} (e_n^2 + f_n^2) \label{eq:Qshunt_def_AC}.
\end{align}
Then the voltage operational limits are defined for every node $n$ as
\begin{equation}
 \underline{V_n}^2 \leq e_n^2 + f_n^2 \leq \overline{V_n}^2, \ \ \forall n \in \mathcal{N},
\end{equation}
and the thermal limits by
\begin{equation}
I_{mn}^2 \leq \overline{I}_{mn}^2, \ \ \forall (m,n) \in \mathcal{L},
\end{equation}
with 
\begin{align}
I_{mn}^2 &= I_{real}^2 + I_{imag}^2\,,\\
I_{real} &= g_{mn} (e_m-e_n) - b_{mn} (f_m-f_n)\,,\\
I_{imag} &= b_{mn} (e_m-e_n) + g_{mn} (f_m-f_n)\,.
\end{align}

\subsubsection{Linearized model}

This model proposed in \cite{bolognani2016existence} approximates linearly \eqref{eq:Pmn_def_AC} and \eqref{eq:Qmn_def_AC}. Note that this approximation does not include the shunt powers, i.e. $P^{shunt}_n = Q^{shunt}_n = 0$, $\forall n \in \mathcal{N}$.
This yields
\begin{align}
P_{mn} &= g_{mn} (e_m - e_n) - b_{mn} (f_m-f_n), \ \ \forall (m,n) \in \mathcal{L} \\
Q_{mn} &= -b_{mn} (e_m - e_n) - g_{mn} (f_m-f_n) , \ \ \forall (m,n) \in \mathcal{L}.
\end{align}
The upper voltage limits and the thermal limits are approximated by a regular polyhedron inscribed in the respective circles of the original limits. The lower voltage operational limit is simply modeled as a lower bound on $e_n$, which means that we make the hypothesis that the angles are small. 
An iterative method could be set up if the approximated solutions are far from feasible solutions of the AC model. However, this turned out to be unnecessary as the decisions taken are most of the time very coherent with those obtained with other models, as illustrated in Section~\ref{sec:test}. This formulation does not account for losses.

\subsubsection{Convex SOCP model} \label{sec:socp}
By introducing variables $u_n \geq 0$, $R_{mn} \geq 0$ and $T_{mn} \in \mathbb{R}$ that substitute the expressions
	$$ \frac{e_n^2 + f_n^2}{\sqrt{2}}, \quad e_m e_n + f_m f_n, \quad f_m e_n - e_m f_n, $$
respectively, constraints \eqref{eq:Pmn_def_AC}-\eqref{eq:Qshunt_def_AC} can be rewritten without $e$ and $f$ as 
\begin{align}
P_{mn} &= g_{mn} \sqrt{2} u_m - g_{mn} R_{mn} - b_{mn} T_{mn} , \ \ \forall (m,n) \in \mathcal{L}\label{eq:Pmn_def_SOCP}\\
Q_{mn} &= - b_{mn} \sqrt{2} u_m + b_{mn} R_{mn} - g_{mn} T_{mn} , \ \ \forall (m,n) \in \mathcal{L}\label{eq:Qmn_def_SOCP}\\
P^{shunt}_n &= g_n^{(sh)} \sqrt{2} u_n, \ \ \forall n \in \mathcal{N} \label{eq:Pshunt_def_SOCP}\\
Q^{shunt}_n &= b_n^{(sh)} \sqrt{2} u_n, \ \ \forall n \in \mathcal{N}. \label{eq:Qshunt_def_SOCP}
\end{align}
The additional set of constraints
\begin{equation}
2 u_m u_n = R_{mn}^2 + T_{mn}^2, \ \ \forall (m,n) \in \mathcal{L}
\end{equation}
are imposed to maintain a relationship between the newly introduced variables.
They are then relaxed to obtain a convex second order cone program:
\begin{equation}
2 u_m u_n \geq R_{mn}^2 + T_{mn}^2 , \ \ \forall (m,n) \in \mathcal{L}. \label{eq:SOCP}
\end{equation}
Voltage limits can be easily rewritten as a function of $u_n$ as 
\begin{equation}
\underline{V_n}^2 \leq \sqrt{2} u_n \leq \overline{V_n}^2, \ \ \forall n \in \mathcal{N}.
\end{equation}
Thermal limits are approximated in the same way as for the non-convex AC model. Note that as $R_{mn} = R_{nm}$ and $T_{mn}=-T_{nm}$, they are in practice replaced by a single variable per branch and constraints \eqref{eq:Pmn_def_SOCP} and \eqref{eq:Qmn_def_SOCP} are updated accordingly.
It is shown in \cite{jabr2006radial} that this relaxation is tight for radial networks under some conditions on the objective function.
These conditions are not met in our formulation since minimizing curtailment is equivalent to maximizing the renewable generation.
To mitigate this issue, the losses term in the objective function must be scaled with a coefficient sufficiently large so that \eqref{eq:SOCP} are tight, but not too large so that the original objective function is still guiding the solution. This tradeoff is further discussed in Section~\ref{sec:test}.

\section{Test instances}
\label{sec:test}
\begin{figure}[h] 
	\centering
	\includegraphics[width=0.3\textwidth]{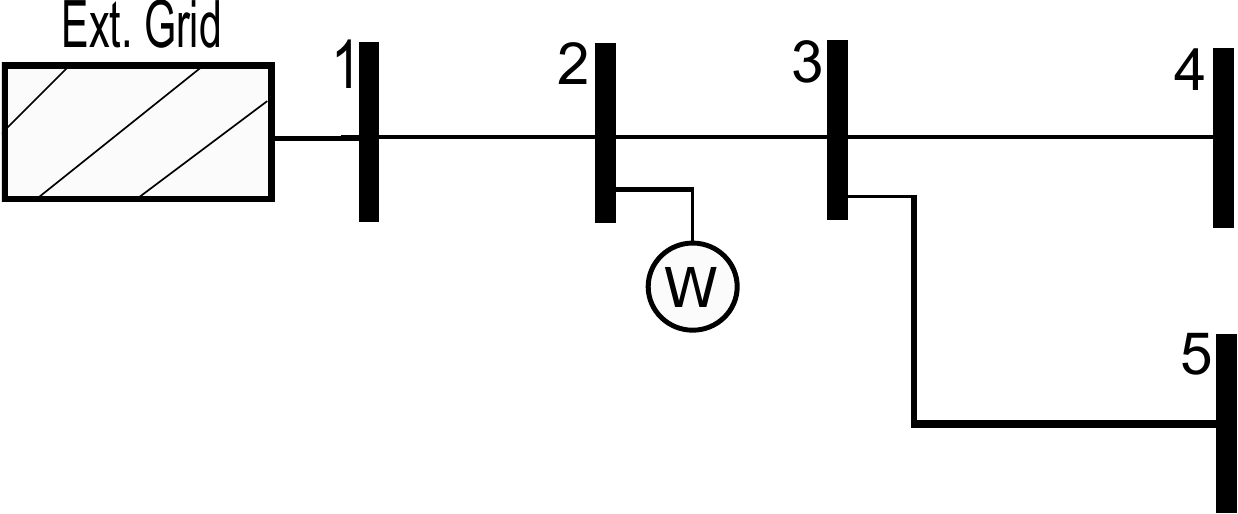}
	\caption{5-bus test system.}
	\label{fig-small}
\end{figure}

\begin{figure}[hbt] 
	\centering
	\includegraphics[width=0.7\textwidth]{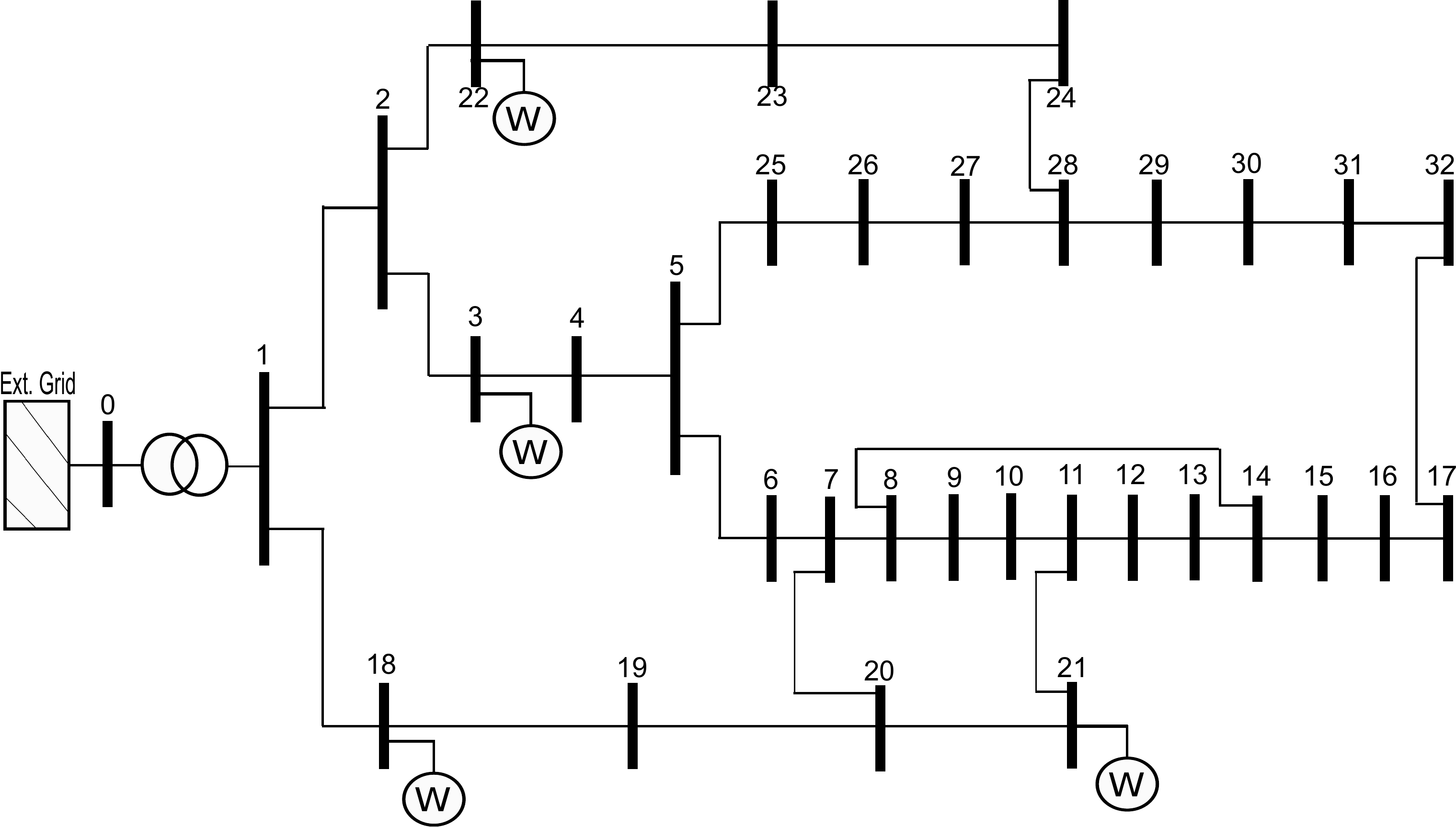}
	\caption{33-bus test system.}
	\label{fig-IEEE34}
\end{figure}

\begin{figure}[hbt] 
	\centering
	\includegraphics[width=0.9\textwidth]{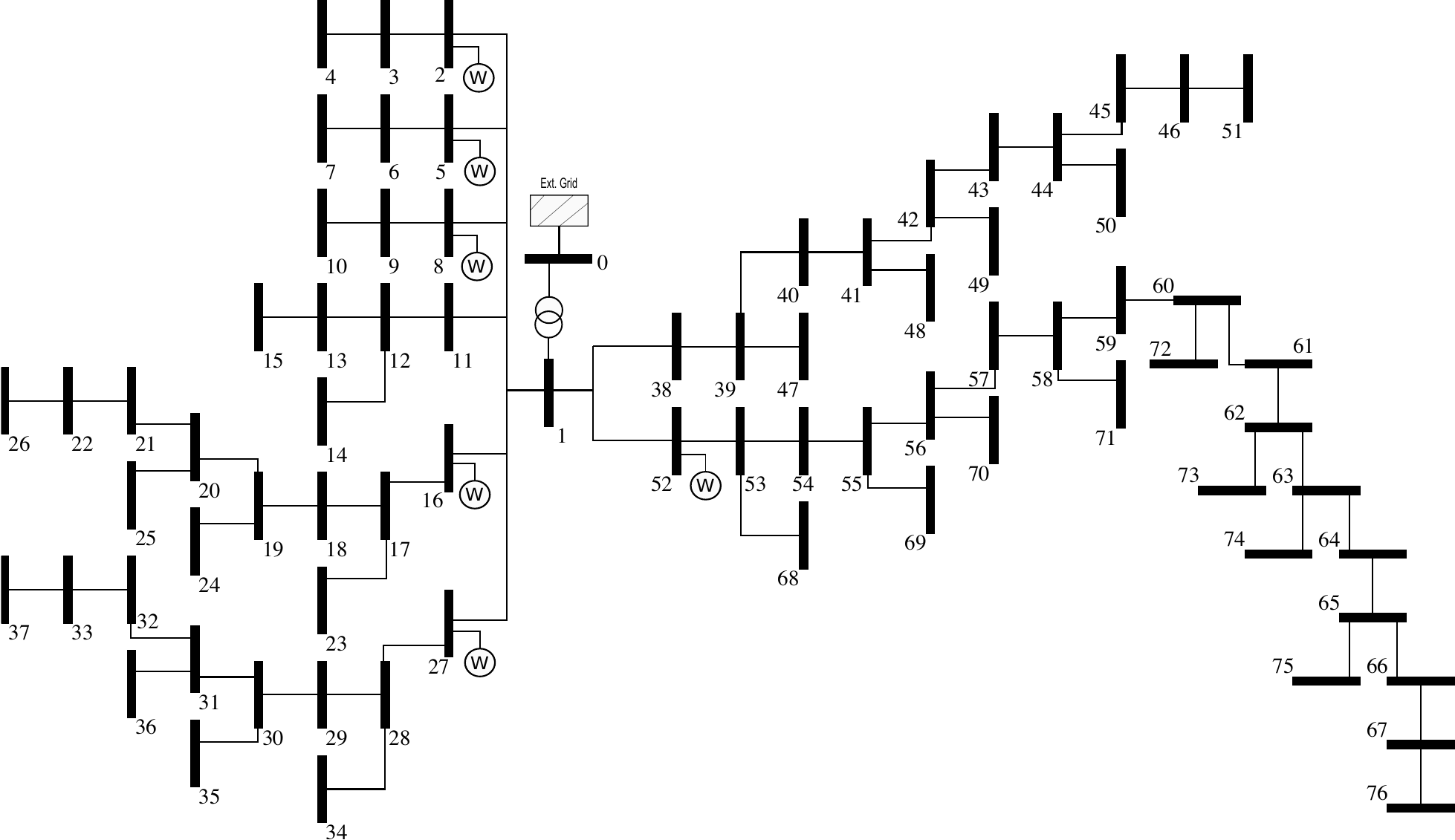}
	\caption{77-bus test system.}
	\label{fig:network}
\end{figure}

We describe below the three test instances of the ANM problem that are used in the results section. The set of models and parameters that are specific to these instances, as well as documentation for their usage, are accessible at \url{http://www.montefiore.ulg.ac.be/~anm/} as Python code. It has been developed to provide a black-box-type simulator that is quick to set up. The DNs on which these instances are based are a toy 5-bus radial test system, a 33-bus non-radial test system \cite{ieee:2015_1}, and a 77-bus radial test system \cite{ukgds}. Table~\ref{tab:case} summarizes some relevant data about these instances. The test systems are also illustrated in Figures~\ref{fig-small}, \ref{fig-IEEE34}, and \ref{fig:network}. The location of the wind generators, which we assume to be curtailable, is indicated by a circled $W$. The 77-bus instance also includes non-curtailable generators that model residential photovoltaic panels.

\begin{table}[hbt]
	\centering
	\scalebox{0.9}{
	\begin{tabular}{|r|r|r|r|r|r|r|r|r|r|}
		\hline
		\textit{case}       & \multicolumn{3}{c|}{\textbf{case5}}                                                         & \multicolumn{3}{c|}{\textbf{case33}}                                                        & \multicolumn{3}{c|}{\textbf{case77}}                                                        \\ \hline
		\textit{flex level}        & \multicolumn{1}{l|}{low} & \multicolumn{1}{l|}{medium} & \multicolumn{1}{l|}{high} & \multicolumn{1}{l|}{low} & \multicolumn{1}{l|}{medium} & \multicolumn{1}{l|}{high} & \multicolumn{1}{l|}{low} & \multicolumn{1}{l|}{medium} & \multicolumn{1}{l|}{high} \\ \hline
		$|\mathcal{N}|$   & \multicolumn{3}{r|}{5}                                                             & \multicolumn{3}{r|}{33}                                                            & \multicolumn{3}{r|}{77}                                                            \\ \hline
		$|\mathcal{L}|$   & \multicolumn{3}{r|}{4}                                                             & \multicolumn{3}{r|}{37}                                                            & \multicolumn{3}{r|}{76}                                                            \\ \hline
		$|\mathcal{G}|$   & \multicolumn{3}{r|}{1}                                                             & \multicolumn{3}{r|}{4}                                                             & \multicolumn{3}{r|}{6 curtailable (out of 59)}                                                            \\ \hline
		$|\mathcal{C}|$   & \multicolumn{3}{r|}{3}                                                             & \multicolumn{3}{r|}{32}                                                            & \multicolumn{3}{r|}{53}                                                            \\ \hline
		$|\mathcal{F}|$   & 1                        & 2                           & 3                         & 11                       & 22                          & 32                        & 11                       & 22                          & 33                        \\ \hline
		\textit{max flex (MW)}          & 0.3                      & 0.6                         & 0.9                       & 0.62                     & 1.3                         & 2                         & 1.71                     & 3.41                        & 5.01                       \\ \hline
		\textit{$\sim$peak load (MW)} & \multicolumn{3}{r|}{11}                                                            & \multicolumn{3}{r|}{9}                                                             & \multicolumn{3}{r|}{18}                                                            \\ \hline
		\begin{tabular}{@{}c@{}}$\mathcal{O}_g\,,$ \\ \textit{$\forall g \in$ wind turbines}\end{tabular} & \multicolumn{3}{c|}{\begin{tabular}{@{}c@{}}$0 \leq P_g \leq 20$\\ $-5 \leq Q_g \leq 5$ \\ $Q_g \leq -0.24 P_g + 6.8$\\ $Q_g \leq 0.24 P_g - 6.8$\end{tabular}}                                                           & \multicolumn{3}{c|}{\begin{tabular}{@{}c@{}}$0 \leq P_g \leq 4.5$\\ $-1 \leq Q_g \leq 1$ \\ $Q_g \leq -0.2 P_g + 1.3$\\ $Q_g \leq 0.2 P_g - 1.3$\end{tabular}}                                                             & \multicolumn{3}{c|}{\begin{tabular}{@{}c@{}}$0 \leq P_g \leq 4.5$\\ $-1 \leq Q_g \leq 1$ \\ $Q_g \leq -0.2 P_g + 1.3$\\ $Q_g \leq 0.2 P_g - 1.3$\end{tabular}}                                                            \\ \hline
	\end{tabular}}
	\caption{Summary of test instances.}
	\label{tab:case}
\end{table}


\begin{figure}[t]
	\centering
	\includegraphics[width=0.85\textwidth]{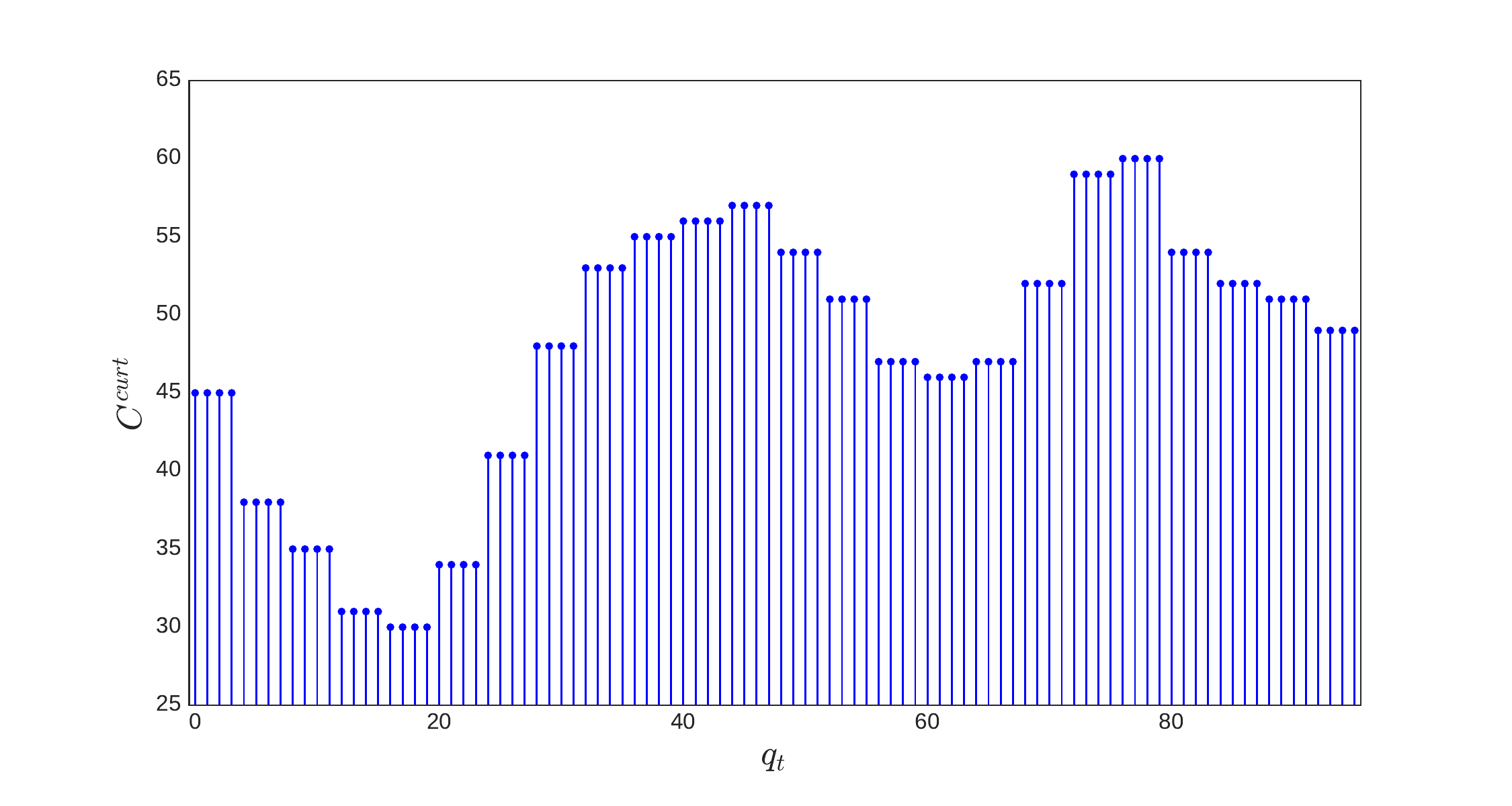}
	\caption{Curtailment prices used in the test instance.}
	\label{fig:curt_costs}
\end{figure}

We consider that the per-unit curtailment prices are the same for all the generators. As described in Section~\ref{sec:mdp_inst}, this price varies through the day and Figure~\ref{fig:curt_costs} specifies the values considered in the test instances. We also use these values for the per-unit cost of the losses, i.e. $C_{loss}(\cdot)$, while the constant $k$ that appear in equation~(\ref{eq:barrier}) is set to $10^4$.
Concerning flexible loads, three different penetration levels exist for each test case. For every configuration, about half of the flexible services offer a downward modulation, followed by an upward rebound effect, and inversely for the other half. The maximal and cumulated modulation magnitude is reported in Table~\ref{tab:case} to illustrate the potential offered by flexible loads in every configuration. Finally, the duration of the modulation signals is of 7 time periods for the 5-bus instances, and from 6 to 24  time periods for the 33 and 77-bus instances.

The conditional density functions defined in equation~(\ref{eq:cond_prob}) are built using a parametric model $\mathcal{P}_{(N,n)}$ that relies on a mixture of $n$ Gaussians to represent the probability distribution $p(w_{t+1} | w_{t},\dots,w_{t-N+1})$ of the next outcome of the process, conditionally to the last $N$ observed outcomes. In particular, the following procedure allows to fit a model $\mathcal{P}_{(N,n)}$ to a set $\{(w^{(i)}_1, w^{(i)}_2, \dots, w^{(i)}_L), i = 1, \dots, I\}$ of time series of normalized realizations of the process of interest:
\begin{enumerate}
\item build a dataset of tuples $(w^{(i)}_{t-N+1}, \dots, w^{(i)}_{t}, w^{(i)}_{t+1}), \forall (i,t) \in \{1,\dots,I\} \times \{N,\dots,L-1\}$;
\item model the joint distribution $p(w_{t-N+1},\dots,w_{t+1})$ of the dataset using a mixture of $n$ Gaussians, by performing a maximum likelihood estimation \cite{redner1984mixture} of the mixture's parameters (i.e. the weight $\eta_i$, mean $\bm{\mu}_i$, and covariance matrix $\bm{\Sigma}_i$ of every Gaussian $i \in \{1,\dots,n\}$);
\item $\forall i \in \{1,\dots,n\}$, deduce from $\bm{\mu}_i$ and $\bm{\Sigma}_i$ the functions $\mu_i^{\cdot|\cdot}(w_{t-N+1}, \dots, w_{t})$ and \linebreak $\sigma_i^{\cdot|\cdot}(w_{t-N+1}, \dots, w_{t})$ that define the mean and standard deviation of $w_{t+1}$, according to the $\text{i}^{\text{th}}$ Gaussian in the mixture and conditional to $w_{t-N+1},\dots,w_{t}$ \cite{bishop2006pattern};
\item produce $p(\cdot | w_{t},\dots,w_{t-N+1})$ as the following mixture of conditional Gaussian distributions: \begin{align}
	p(\cdot | w_{t},\dots,w_{t-N+1}) = \sum_{i=1}^n \eta_i \mathcal{N}\left(\mu_i^{\cdot|\cdot}(w_{t-N+1}, \dots, w_{t}), \sigma_i^{\cdot|\cdot}(w_{t-N+1}, \dots, w_{t})\right)\,.
\end{align}
\end{enumerate}
In order to determine an adequate value of the model's hyper-parameters $n$ and $N$ for each process, we relied on an Approximate Bayesian Computation (ABC) method \cite{marin2012approximate}. Such an approach consists in sampling trajectories from each model and to compare them with the original data to estimate its posterior probability among the set $\Theta$ of candidate models \cite{grelaud2009abc}.
Using $\Theta =\{\mathcal{P}_{(N,m)} |{\scriptstyle n \in \{1,\dots,20\},N \in \{1,\dots,3\}}\}$, the most likely parameters identified by this model choice technique are presented in Table~\ref{tab:models}. We refer the interested reader to \cite{gemine16gaussian} for more details on the modeling approach.
\begin{table}[t]
   \centering
   \caption{\label{tab:models} Parameters of the stochastic models used in the implementation of the benchmark.}
   \begin{tabularx}{\textwidth}{X r r}
     	\hline
	 & $N$ & $n$ \\
	\hline
	Wind speed	($N_v$)	&	1	&	1	\\
	Solar irradiance ($N_{ir}$)	&	1	&	10	\\
	Load consumption ($N_{loads}$)	&	2	&	10	\\
	\hline
   \end{tabularx}

\end{table}

The datasets that we used are real measurements of the wind speed\footnote{\url{http://www.nrel.gov/electricity/transmission/eastern_wind_dataset.html}} and of the solar irradiance\footnote{\url{http://solargis.info/}}. For the residential consumption data, a single stochastic model has been learned from measurements of a Belgian distribution network and it is used for all the loads of the test instance. However, this model differs among the loads through the use of a scaling factor. The implementation of the statistical algorithms relies on both SciPy \cite{jones2014scipy} and Scikit-learn \cite{pedregosa2011scikit}, two Python libraries.

\section{Numerical results}
\label{sec:num_results}

The goals of this section are to illustrate the operational planning problem and the test instances, as well as to provide some empirical evaluations of the proposed lookahead policy for the considered network models and for scenario trees of varying complexity. In particular, the policy $\hat{\pi}^*_{\hat{\mathcal{M}}_t}(\bm{s}_t)$ defined by problem (\ref{eq:msp1})-(\ref{eq:msp2}) was applied to every test instance and penetration level of the flexible loads. The empirical expected return of the policy, for a given test instance, level of flexibility, network model, and scenario tree complexity, is determined from $50$ runs of $288$ time steps (i.e. of 3 days), each run $i$ corresponding to the following sequence:
\begin{enumerate}
\item \label{it:init} Initialize the state vector $\bm{s}_0$ by setting all the flexible loads as inactive and by sampling stochastic components from the joint distributions learned when building the test instance.
\item Run a simulation of 288 time steps, where, at every time step, problem (\ref{eq:msp1})-(\ref{eq:msp2}) is implemented as follow:
\begin{enumerate}
\item sample $100$ trajectories of the exogenous variables over a lookahead horizon of length $T = 10$, i.e. trajectories $(\bm{w}^{(j)}_t,\dots,\bm{w}^{(j)}_{t+9})$, with $j = 1,\dots,100$;
\item determine the corresponding trajectories of the potential (i.e. not accounting for modulation instructions) power injections of the devices, as they are fully determined by the current state $\bm{s}_t$ and by $(\bm{w}^{(j)}_t, \dots, \bm{w}^{(j)}_{t+9})$;
\item cluster the $100$ trajectories of power injections into $W$ scenarios, using a hierarchical clustering method and Ward's distance \cite{hastie2005elements};
\item build the corresponding clusters of outcome trajectories, i.e. \[\tilde{\mathcal{W}}^T_t = \{(\tilde{\bm{w}}^{(k)}_{t}, \dots, \tilde{\bm{w}}^{(k)}_{t+9}) | k =1,\dots,W\}\,,\] where $\tilde{\bm{w}}^{(k)}_{t'}$ denotes the centroid of cluster $k$ at time $t' \in \{t,\dots,t+9\}$, and compute the probabilities $\mathbb{P}_k$ of the resulting scenarios as \[{\mathbb{P}_k = \frac{\text{number of trajectories in cluster }k}{100}}\,;\]
\item solve problem $\hat{\mathcal{M}}_t$ with a discount factor $\gamma = 0.99$ and over the scenario tree defined by outcomes of $\tilde{\mathcal{W}}^T_t$ and probabilities $(\mathbb{P}_1,\dots,\mathbb{P}_W)$;
\item recover the action vector $\bm{a}_t$ to apply to the system.
\end{enumerate}
\end{enumerate}

The motivation behind the use of Ward's method to cluster trajectories is that it is a minimum variance method, which means that the trajectories of a cluster were selected because they are close to its centroid, in comparison to trajectories of other clusters. Consequently, the scenarios used in the optimization model, which are the centroids of the clusters, differ minimally from the trajectories it summarizes.

The implementation has been done using the Python code mentioned in Section~\ref{sec:test} to simulate the system and Pyomo \cite{hart2012pyomo} to build the mathematical programs. These programs were solved by BONMIN \cite{bonmin} in the MINLP case, and by Gurobi in the MISOCP and MILP cases. At each time step, a budget of 10 minutes is allowed to solve the mathematical program. If the solver reaches the time limit, the current best solution is applied to the system if a feasible solution is available, or the whole simulation run fails if no solution was found. Both solvers stop before the time limit if they reach a relative optimality gap of 1\%. Note that BONMIN performs local optimization and must be seen as an heuristic method to solve the non-convex MINLPs, as it comes with no optimality guarantees. In the MISOCP case, the scaling factor of the losses discussed in Section~\ref{sec:socp} was fixed empirically to 3.  For every combination of test instance, level of flexibility, and network model, the same runs were performed with a scenario tree $\tilde{\mathcal{W}}^T_t$ of one scenario (i.e. the mean of the sampled trajectories) and of three scenarios. A version of the problem with perfect information, i.e. with a scenario tree consisting of the actual future trajectory of the exogenous information, was also simulated to obtain a reference value of performance. The overall simulation was carried on in a high-performance computing environment with 128 cores. Each run being limited to a single core, such an infrastructure enabled hundreds of simulations to run in parallel and thus to speed up computations by the same factor. Ignoring failed simulation runs, more than 1 million of mathematical programs were solved for a cumulated time budget of approximately 1122 days.

The empirical estimations of the expected return reported in the following results are computed as:
\begin{align}
\underset{\bm{s} \sim p_0(\cdot)}{\mathbb{E}}\left\{J^{\hat{\pi}^*}(\bm{s}) \right\} \approx \frac{1}{50}  \sum_{i=1}^{50} \sum_{t=0}^{287} 0.99^{t} r^{(i)}_t\,,
\end{align}
where $r^{(i)}_{t}$ corresponds to the instantaneous reward observed during the $n^{\text{th}}$ simulation run at time step $t$, and where $p_0(\cdot)$ denotes the probability distribution described at step (\ref{it:init}). Tables~\ref{fig:case5_summary}, \ref{fig:case33_summary}, and \ref{fig:case75_summary}, summarize the results of the simulation runs for the 5-bus, 33-bus, and 77-bus test systems, respectively. The first columns identifies the test instance configuration and the two latter columns report the expected return and the distribution of solution time. The blue and red bars denote the contributions to the expected return of the expected costs (including losses) and of the penalties from constraint violations, respectively. The box plots' whiskers cover the whole range of the observed solution times and the red makers indicate the median time.

We can first observe from the simulation results of the 5-bus test system that having a perfect forecast of the evolution of the system yields significantly better returns than when decisions are subject to uncertainty. It also shows that considering three possible future scenarios can significantly improve over an optimization performed on the average future scenario, to the expense of the solution time. Among the three network models, the SOCP relaxation is the one that induces the most penalties. These penalties also appear in the deterministic case, which implies that the relaxation is not always tight. We also observe that the policy slightly benefits from an increase of the flexibility level of loads in the deterministic setting but not in the presence of uncertainty. Given the small size of the 5-bus test system, the solution times are very good for every configuration with the exception of the MINLPs, for which worst-case solution time already reaches the time limit for all lookahead models.

The results for the 33-bus test system are quite similar, with the notable exception of the solution times. Both the SOCP and LP network models still produce mathematical programs that can be solved within a reasonable time budget. On the other hand, the time limit is very often reached when solving the MINLPs\footnote{Reported solution time can be larger than the time limit. It happens when the solver is executing a complex routine for some amount time before being able to check the limit.}. The simulation runs even failed in the two most complex configurations as the solver could not find any feasible solution within the time limit. This observation can be extended to the simulation results of the 77-bus test system, with the difference that, in the 1-scenario configurations, the mean and median solution times are lower while the worst-case time is even larger. None of the 3-scenario runs succeeded for the NLP network model and even the SOCP model produces significantly increased solution times, with worst cases reaching the time limit.
\begin{table}[ht]
	\centering
	\includegraphics[width=1.0\textwidth]{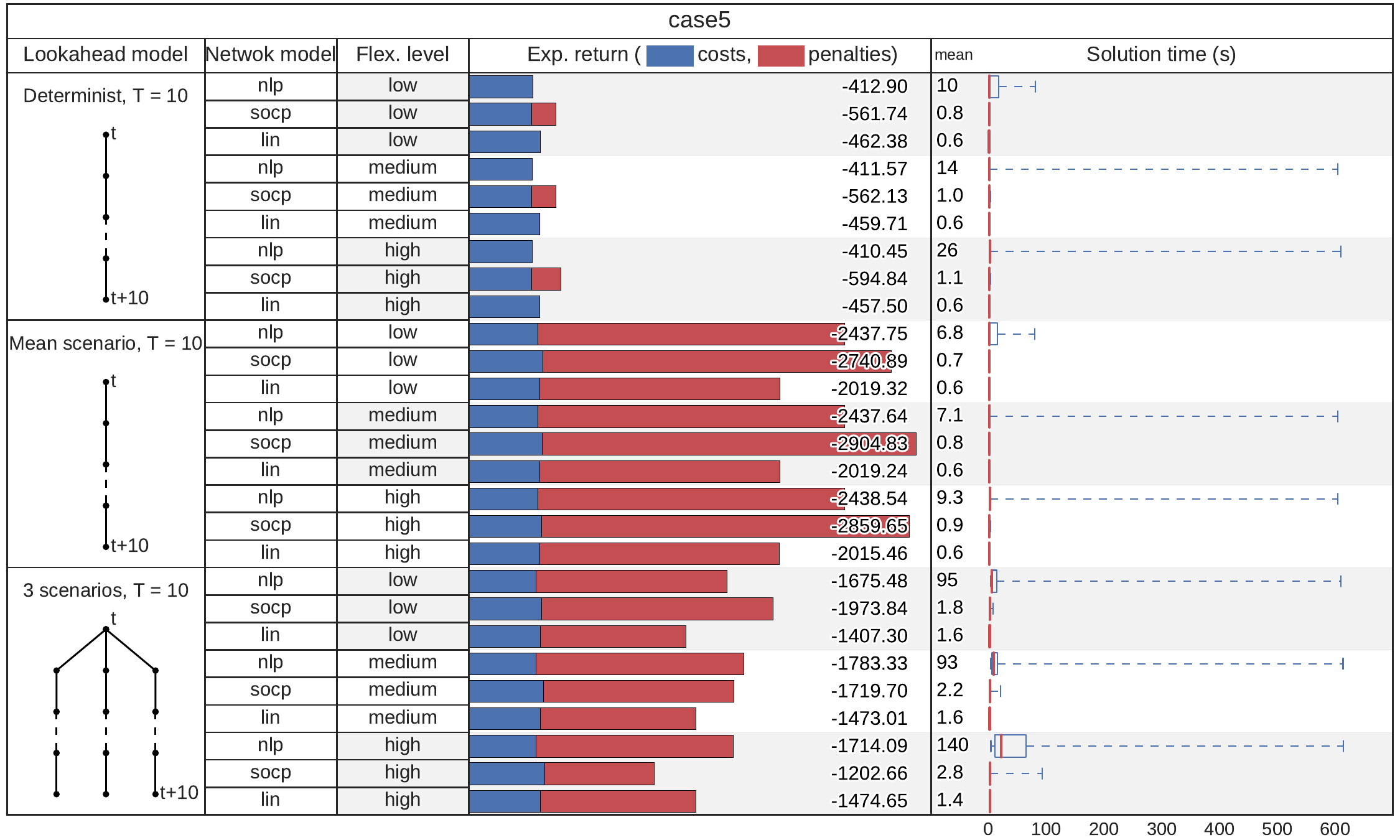}
	\caption{Estimation of expected return and distribution of solver time for the 5-bus test system.}
	\label{fig:case5_summary}
\end{table}
\begin{table}[ht]
	\centering
	\includegraphics[width=1.0\textwidth]{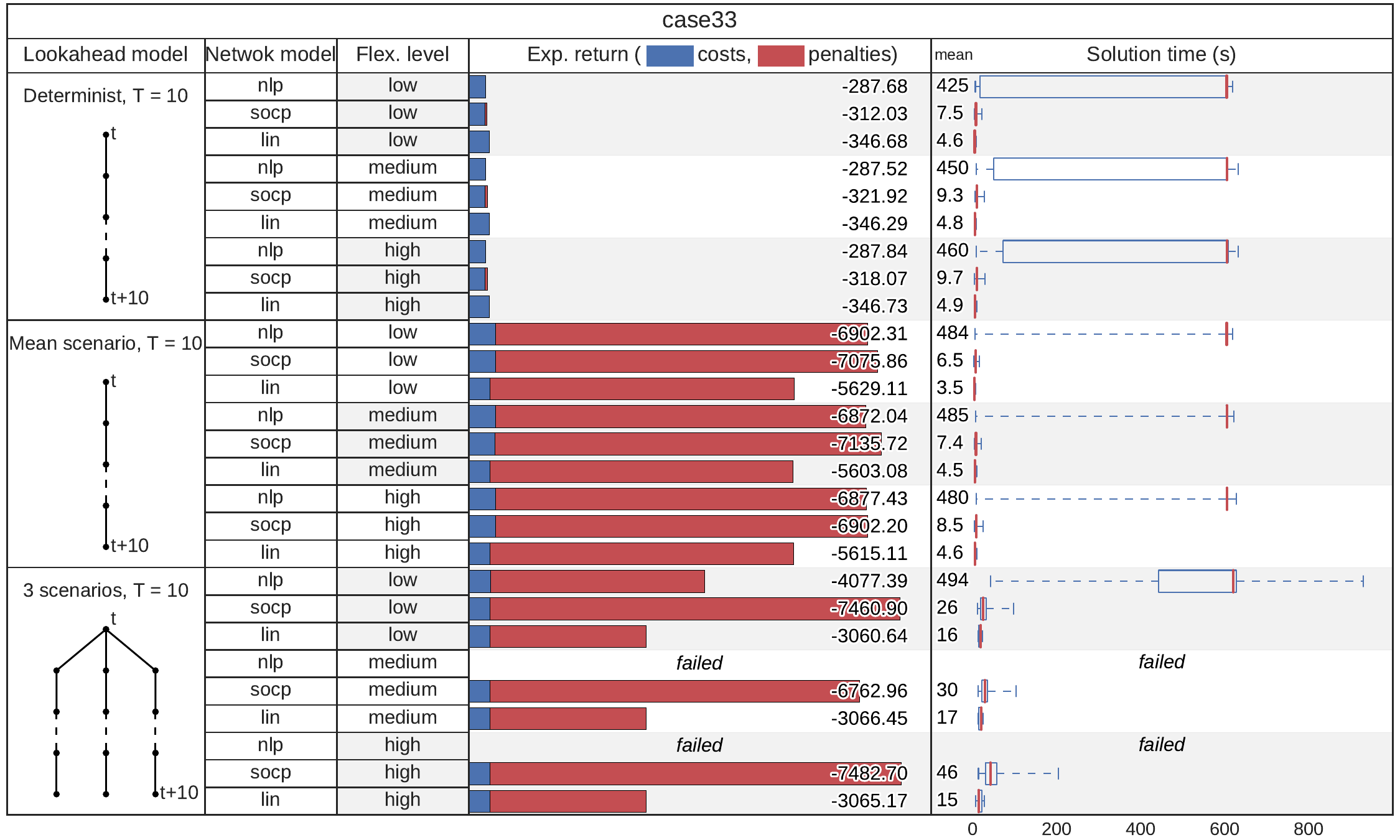}
	\caption{Estimation of expected return and distribution of solver time for the 33-bus test system.}
	\label{fig:case33_summary}
\end{table}	
\begin{table}[ht]
	\centering
	\includegraphics[width=1.0\textwidth]{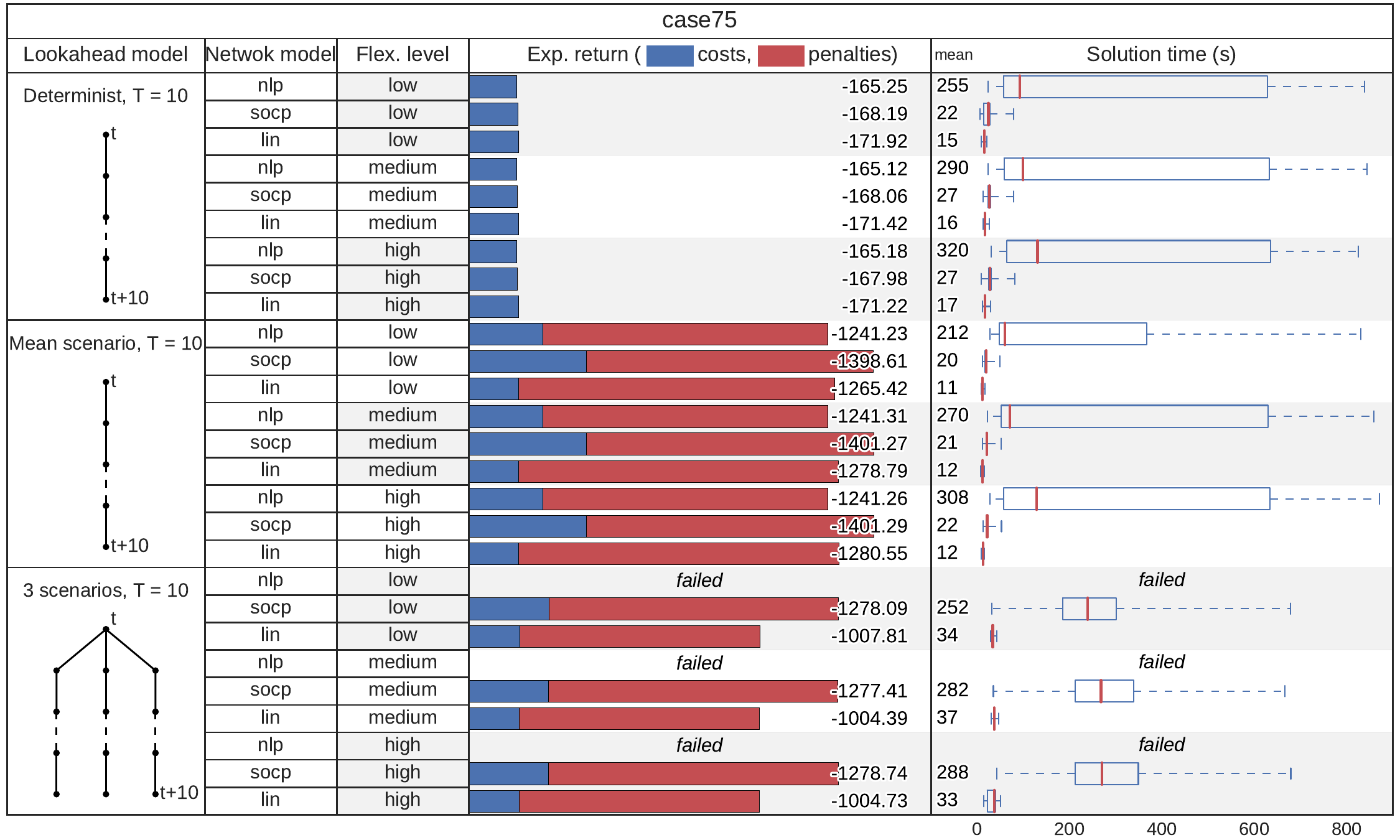}
	\caption{Estimation of expected return and distribution of solver time for the 77-bus test system.}
	\label{fig:case75_summary}
\end{table}

\begin{figure}[ht]
	\captionsetup[subfigure]{labelformat=empty, captionskip=3pt}
	\centering
	\setbox1=\hbox{\includegraphics[width=.45\textwidth]{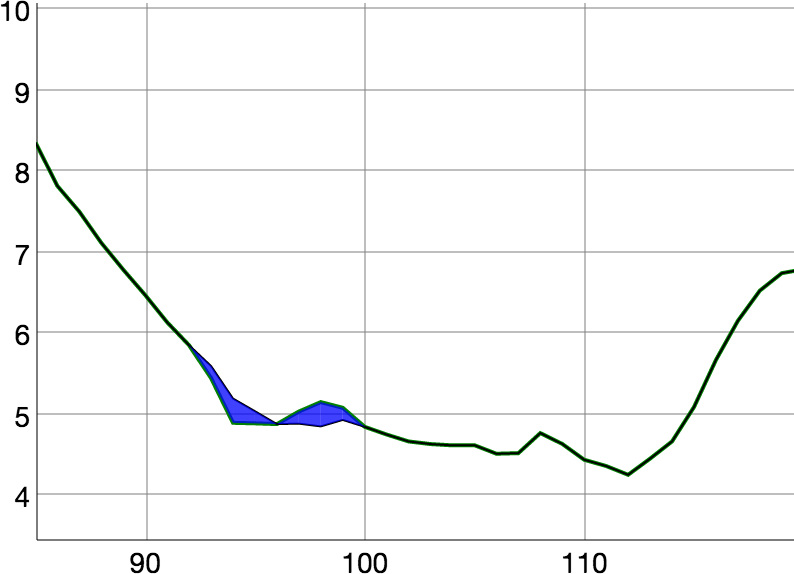}}
	\raisebox{\dimexpr.5\ht1-.5\height}{\begin{minipage}{0.5em}
		\rotatebox{90}{{Prod. (MW)}}
	\end{minipage}}
	\subfloat[Time step]{\includegraphics[width=.45\textwidth]{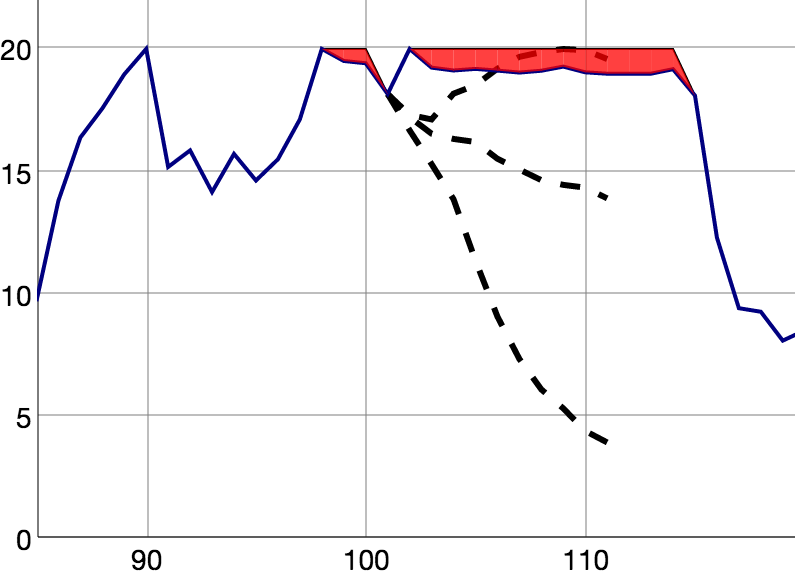}}
	~
	\raisebox{\dimexpr.5\ht1-.5\height}{\begin{minipage}{0.5em}
		\rotatebox{90}{{Cons. (MW)}}
	\end{minipage}}
	\subfloat[Time step]{\includegraphics[width=.45\textwidth]{screenshots/case5_nlp_cons.png}}\\
	\setbox1=\hbox{\includegraphics[width=.45\textwidth]{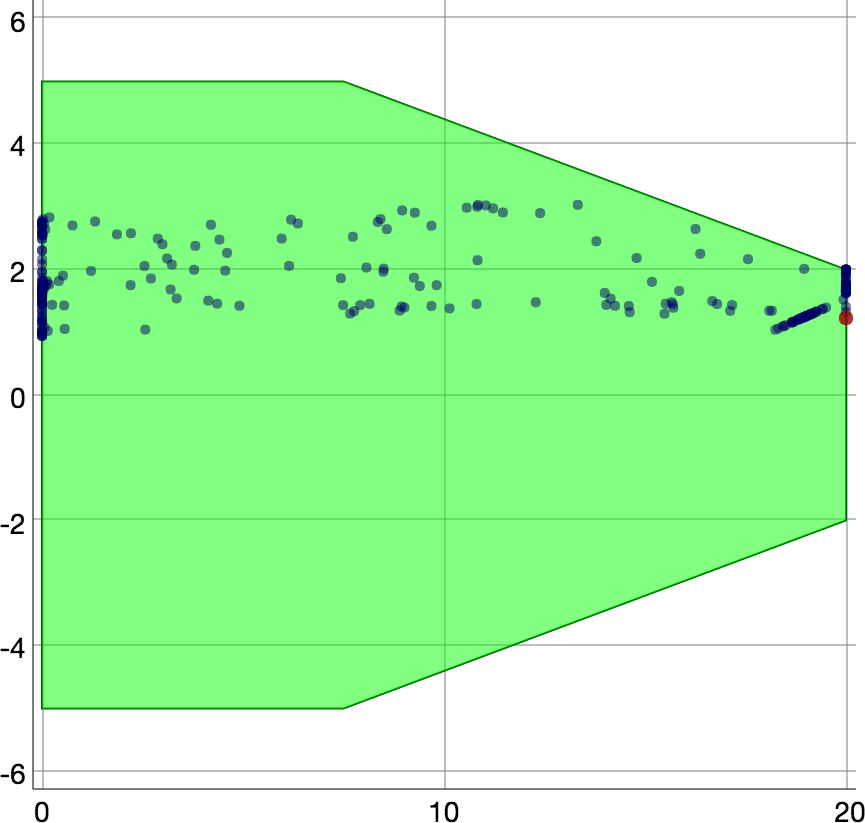}}
	\raisebox{\dimexpr.5\ht1-.5\height}{\begin{minipage}{0.6em}
		\rotatebox{90}{{$\mathrm{|I|/\bar{I}}$}}
	\end{minipage}}
	\subfloat[]{\raisebox{\dimexpr.5\ht1-.5\height}{\stackunder[3pt]{\includegraphics[width=.48\textwidth]{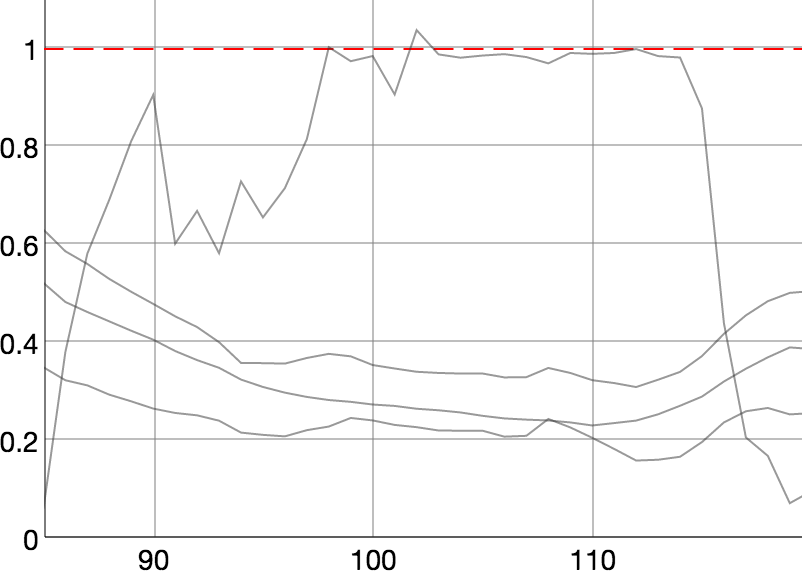}}{Time step}}}
	~
	\raisebox{\dimexpr.5\ht1-.5\height}{\begin{minipage}{0.6em}
			\rotatebox{90}{{Q (MW)}}
		\end{minipage}}
		\subfloat[P (MW)]{\includegraphics[width=.45\textwidth]{screenshots/case5_nlp_dev.png}}
	\caption{Illustrations of part of a 5-bus and low-flexibility simulation run, with a 3-scenarios lookahead model and a NLP network model.}
	\label{fig:case5nlp}
\end{figure}

\begin{figure}[ht]
	\captionsetup[subfigure]{labelformat=empty, captionskip=3pt}
	\centering
	\setbox1=\hbox{\includegraphics[width=.45\textwidth]{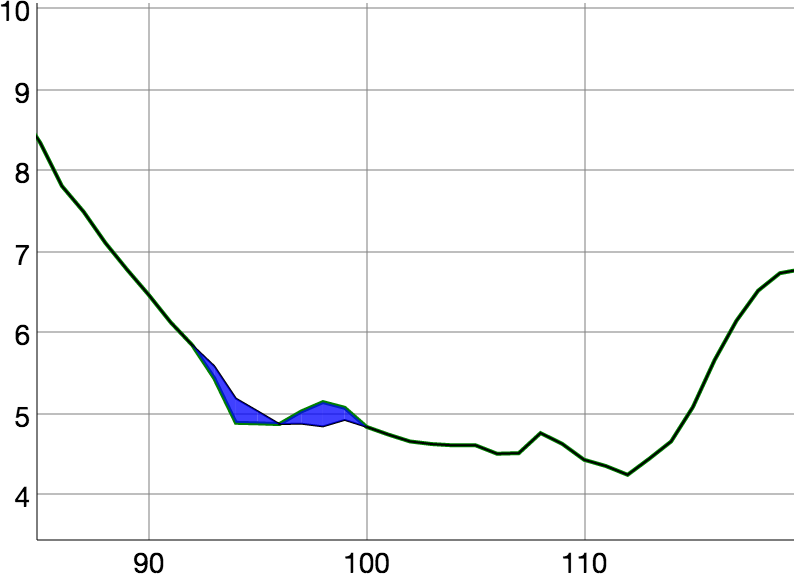}}
	\raisebox{\dimexpr.5\ht1-.5\height}{\begin{minipage}{0.5em}
			\rotatebox{90}{{Prod. (MW)}}
		\end{minipage}}
	\subfloat[Time step]{\includegraphics[width=.45\textwidth]{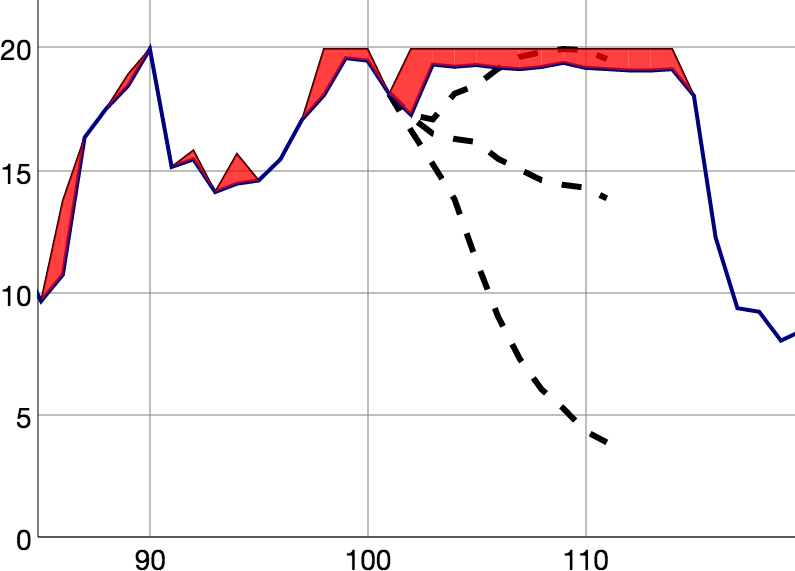}}
	~
	\raisebox{\dimexpr.5\ht1-.5\height}{\begin{minipage}{0.5em}
			\rotatebox{90}{{Cons. (MW)}}
	\end{minipage}}
	\subfloat[Time step]{\includegraphics[width=.45\textwidth]{screenshots/case5_socp_cons.png}}\\
	\setbox1=\hbox{\includegraphics[width=.45\textwidth]{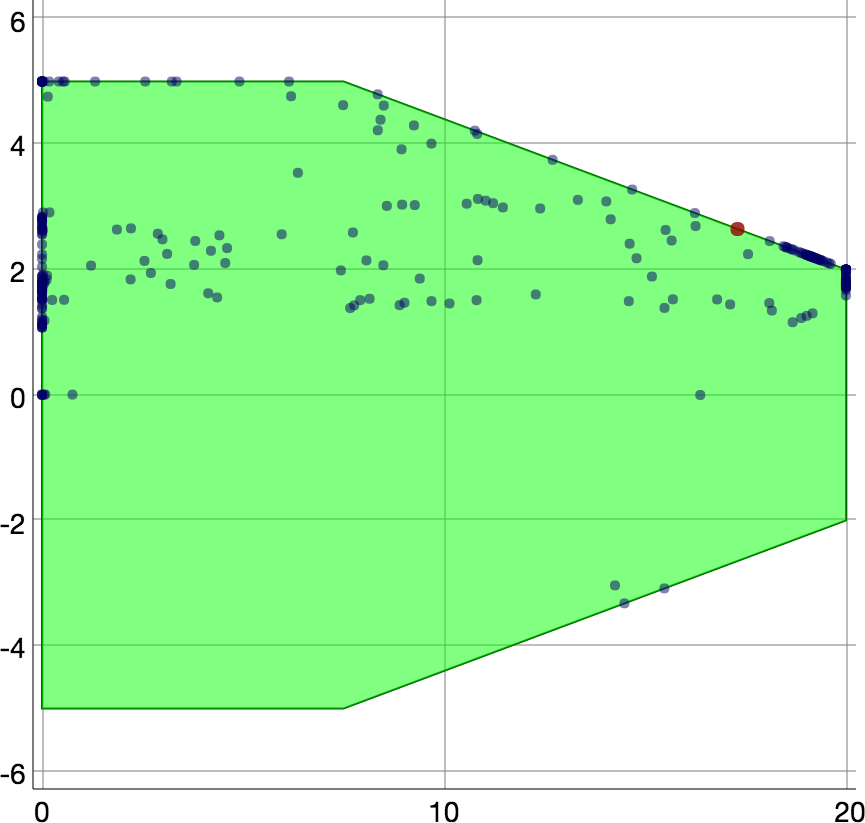}}
	\raisebox{\dimexpr.5\ht1-.5\height}{\begin{minipage}{0.6em}
		\rotatebox{90}{{$\mathrm{|I|/\bar{I}}$}}
	\end{minipage}}
	\subfloat[]{\raisebox{\dimexpr.5\ht1-.5\height}{\stackunder[3pt]{\includegraphics[width=.48\textwidth]{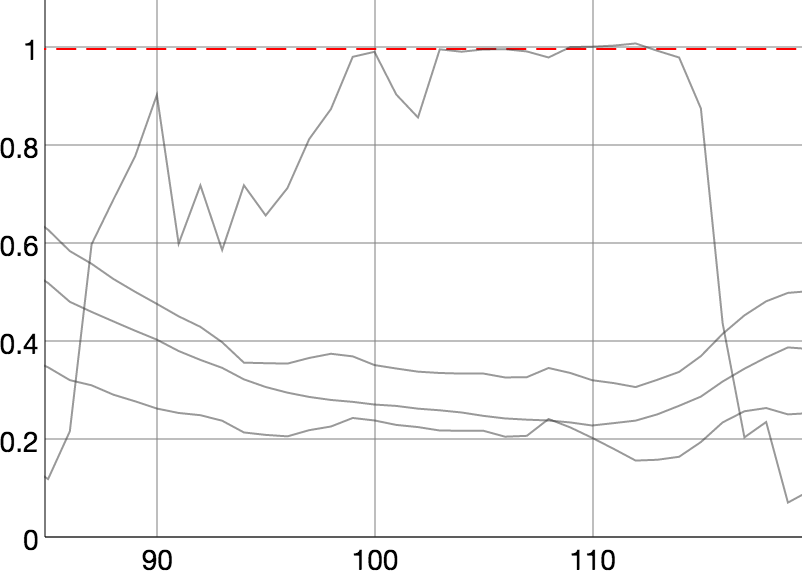}}{Time step}}}
	~
	\raisebox{\dimexpr.5\ht1-.5\height}{\begin{minipage}{0.6em}
		\rotatebox{90}{{Q (MW)}}
	\end{minipage}}
	\subfloat[P (MW)]{\includegraphics[width=.45\textwidth]{screenshots/case5_socp_dev.png}}
	\caption{Illustrations of part of a 5-bus and low-flexibility simulation run, with a 3-scenarios lookahead model and a SOCP network model.}
	\label{fig:case5socp}
\end{figure}

\begin{figure}[ht]
	\captionsetup[subfigure]{labelformat=empty, captionskip=3pt}
	\centering
	\setbox1=\hbox{\includegraphics[width=.45\textwidth]{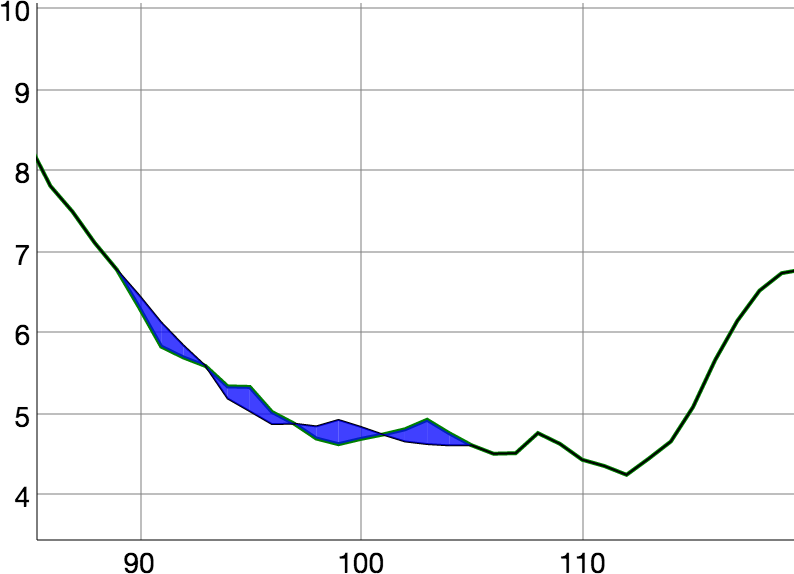}}
	\raisebox{\dimexpr.5\ht1-.5\height}{\begin{minipage}{0.5em}
			\rotatebox{90}{{Prod. (MW)}}
		\end{minipage}}
	\subfloat[Time step]{\includegraphics[width=.45\textwidth]{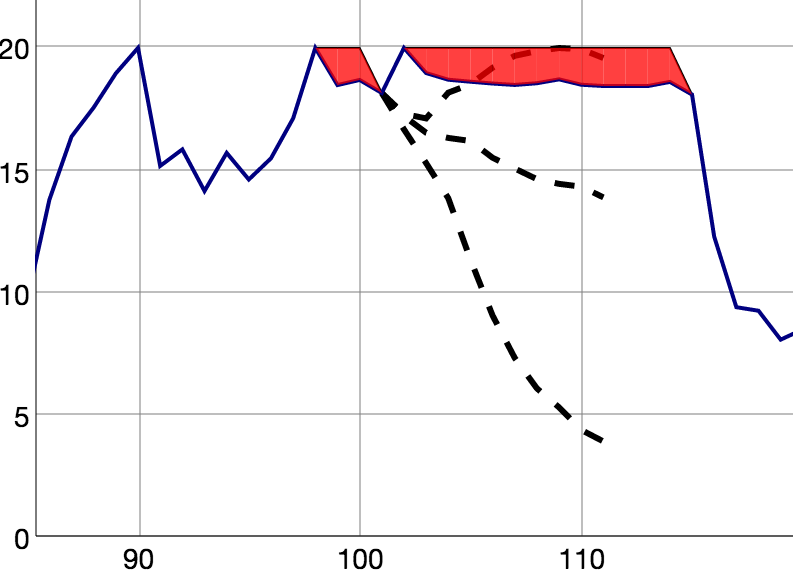}}
	~
	\raisebox{\dimexpr.5\ht1-.5\height}{\begin{minipage}{0.5em}
			\rotatebox{90}{{Cons. (MW)}}
	\end{minipage}}
	\subfloat[Time step]{\includegraphics[width=.45\textwidth]{screenshots/case5_lin_cons.png}}\\
	\setbox1=\hbox{\includegraphics[width=.45\textwidth]{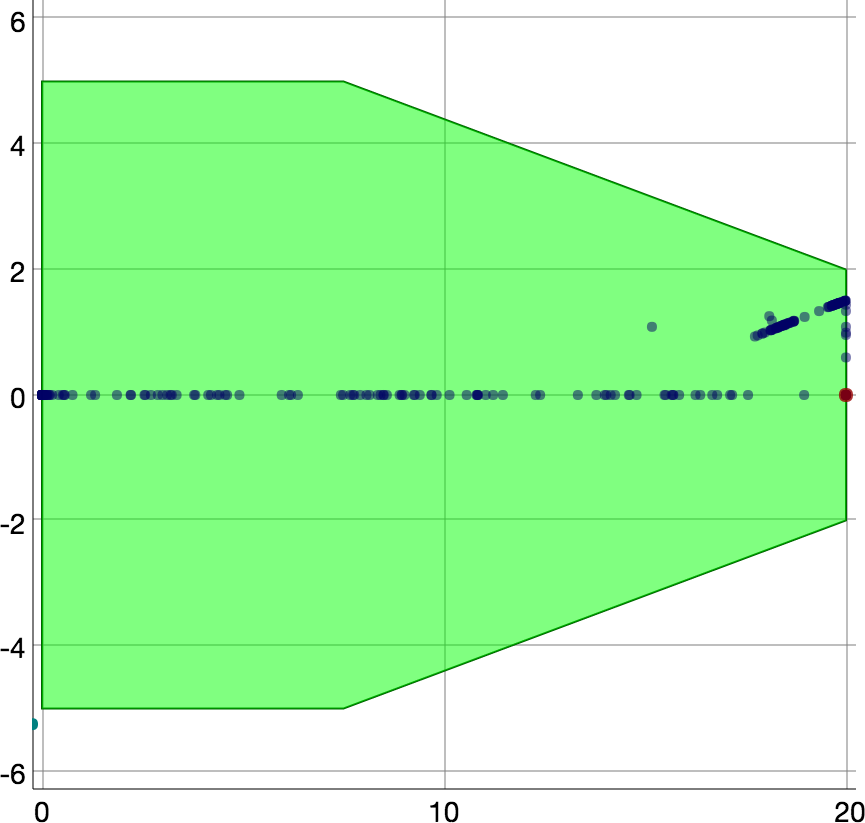}}
	\raisebox{\dimexpr.5\ht1-.5\height}{\begin{minipage}{0.6em}
		\rotatebox{90}{{$\mathrm{|I|/\bar{I}}$}}
	\end{minipage}}
	\subfloat[]{\raisebox{\dimexpr.5\ht1-.5\height}{\stackunder[3pt]{\includegraphics[width=.48\textwidth]{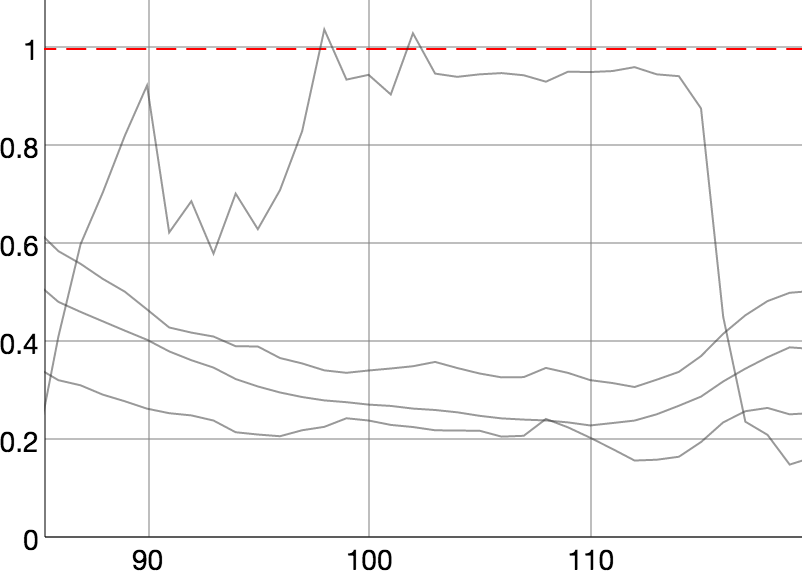}}{Time step}}}
	~
	\raisebox{\dimexpr.5\ht1-.5\height}{\begin{minipage}{0.6em}
		\rotatebox{90}{{Q (MW)}}
	\end{minipage}}
	\subfloat[P (MW)]{\includegraphics[width=.45\textwidth]{screenshots/case5_lin_dev.png}}
	\caption{Illustrations of part of a 5-bus and low-flexibility simulation run, with a 3-scenarios lookahead model and a linear network model.}
	\label{fig:case5lin}
\end{figure}

Finally, a part of a 5-bus and low-flexibility simulation run is illustrated in Figures~\ref{fig:case5nlp}, \ref{fig:case5socp}, and \ref{fig:case5lin}, with a 3-scenarios lookahead model and a NLP, SOCP, and linear network model, respectively. The dashed lines in the upper-left subplots represent the estimated production in the 3 scenarios of the lookahead model at time step 101 (i.e. when computing decisions for time step 102 and onwards). The bottom-right subplots represent the generator's P-Q operating points for the whole simulation and the red point corresponds to time step 102. Notable differences can be observed among the network models. At time step 102, both the NLP and LP approaches show a violation of a thermal constraint because of an inadequate scenario tree, but the SOCP model is, on the contrary, quite conservative. This behavior is likely due to the scaling of the losses term in this latter model, as suggested by its chart of P-Q set-points. The policy did not explicitly computed a curtailment of active power but chose an aggressive Q set-point, which led to a power curtailment due to the P-Q capabilities of the generator. This phenomenon is observed several times in the reported simulation, in particular at and prior to time step 102. The NLP and LP models show a similar curtailment pattern, with the latter inducing more curtailment and an over-satisfaction of the thermal limit. We suspect the cause to be the non-inclusion of the losses in the LP model, which may also explain why this approach activates more flexible services than the two other approaches. Another consequence of not accounting for the losses is that the policy makes little use of the generator's reactive capabilities, as shown by the lower-right subplot of Figure~\ref{fig:case5lin}. In accordance with results of Tables~\ref{fig:case5_summary}, \ref{fig:case33_summary}, and \ref{fig:case75_summary}, the SOCP network model is not always tight and, around time step 110, shows constraint violations in the lower-left subplot of Figure~\ref{fig:case5socp}, while the two other network models keep the system within the operational limits.

\clearpage

\section{Conclusions}
\label{sec:conclusion}

Active Network Management is an alternative or a complement to network reinforcement in case of massive integration of renewable energy in distribution systems in the future.
Mathematically, operational planning, which is the preventive version of active network management we consider in this paper, is an optimal sequential decision-making problem under uncertainty.  The properties of the operational planning problem that we want to highlight are the need to optimize over a sufficiently long time horizon, to account for uncertainty of generation and consumption, and to model the discrete decisions related to the activation of flexibility services.
In an attempt not to restrict ourselves to one solution method and one research community, we provide a formulation of this problem as a Markov Decision Process (MDP), which does not call for a particular solution method.
We provide a simulator and several test beds at \url{http://www.montefiore.ulg.ac.be/~anm/} along with this formulation to foster research in this field, and ease future comparison of results. Although these benchmarks are not taken from real systems, their properties are coherent with what system operators could face in real life.
We detail one possible solution method, which is a lookahead optimization model, then cast the MDP as a sequence of MINLPs, MISOCPs, or MILPs, and provide results on the benchmarks we created. Results show that state of the art open source local solvers for MINLP can show good performance on the test instances of limited size, at least when we approximate the stochastic program with few scenarios. Solving the MISOCPs and MILPs is however much more tractable, to the expense of the network model accuracy. In particular, the results of the MILP approximation suggest that it could scale to larger test systems and scenario trees.
On the modeling side, we considered that all buses except the slack bus are P-Q buses, and that the power factors of the loads are constant while the generators are flexible as defined by their P-Q capabilities. Possible extensions of this work could be to consider the control of steerable synchronous generation, and of generators with time coupling constraints (e.g. combined heat and power generation).
As mentioned in Section~\ref{sec:operationalPlanning}, other approaches exist to control the system, such as modulating the tariff signal(s), acting on the topology of the network, or using distributed storage sources. We did not model either the automatic regulation devices that often exist in distribution systems, such as On Load Tap Changers (OLTCs) of transformers that automatically adapt to control the voltage level.  These automatic regulation devices have been recently addressed in \cite{macedo2015optimal}. We believe that all of these aspects should be considered in a real life solution.
However, computational experiments show that we are at the limit of what can be achieved with modern computers and standard mathematical programing tools. Furthermore, including a more detailed representation of the physical system makes the problem yet more discrete (OLTCs), and more uncertain (for instance, if flexibility services are not as well characterized as what we have assumed). Our experiments also show that increasing the number of scenarios, or stages of the stochastic program, would probably significantly improve the policies.  All these observations suggest further research for tailored approximation or decomposition techniques, for instance techniques relying on the dynamic programming framework, in particular \emph{approximate dynamic programming}, or simulation methods, such as \emph{direct policy search} \cite{busoniu2011cross} or \emph{Monte-Carlo tree search} \cite{bertsekas1996neuro,busoniu2010reinforcement}, or other approaches from the robust and stochastic programming community \cite{powell2014clearing}.  Actually the benchmarks that we proposed makes the comparison of new techniques possible.

\section*{Acknowledgements}
This research is supported by the public service of Wallonia - Department of Energy and Sustainable Building within the framework of the GREDOR project. The authors give their thanks for the financial support of the Belgian Network DYSCO, an Inter-university Attraction Poles Program initiated by the Belgian State, Science Policy Office.
The authors would also like to thank Rapha\"el Fonteneau for his precious advices and comments.

\bibliographystyle{unsrt}
\bibliography{gemine2016active}

\begin{thebibliography}{10}

\bibitem{res_support}
D.~Fouquet and T.B. Johansson.
\newblock European renewable energy policy at crossroads -- focus on
  electricity support mechanisms.
\newblock {\em Energy Policy}, 36(11):4079--4092, 2008.

\bibitem{cornelusse2015global}
Bertrand Corn{\'e}lusse, David Vangulick, Mevludin Glavic, and Damien Ernst.
\newblock Global capacity announcement of electrical distribution systems: A
  pragmatic approach.
\newblock {\em Sustainable Energy, Grids and Networks}, 4:43--53, 2015.

\bibitem{wang2010dg}
David~TC Wang, Luis~F Ochoa, and Gareth~P Harrison.
\newblock Dg impact on investment deferral: Network planning and security of
  supply.
\newblock {\em Power Systems, IEEE Transactions on}, 25(2):1134--1141, 2010.

\bibitem{joseph_et_al}
J.A.P. Lopes, N.~Hatziargyriou, J.~Mutale, P.~Djapic, and N.~Jenkins.
\newblock Integrating distributed generation into electric power systems: A
  review of drivers, challenges and opportunities.
\newblock {\em Electric Power Systems Research}, 77(9):1189--1203, 2007.

\bibitem{liew_goran}
S.N. Liew and G.~Strbac.
\newblock Maximising penetration of wind generation in existing distribution
  networks.
\newblock {\em IET Generation Transmission and Distribution}, 149(3):256--262,
  2002.

\bibitem{nando_capacity}
L.F. Ochoa, C.J. Dent, and G.P. Harrison.
\newblock Distribution network capacity assessment: Variable {DG} and active
  networks.
\newblock {\em IEEE Transactions on Power Systems}, 25(1):87--95, 2010.

\bibitem{dolan}
M.~J. Dolan, E.~M. Davidson, I.~Kockar, G.~W. Ault, and S.~D.~J. McArthur.
\newblock Distribution power flow management utilizing an online optimal power
  flow technique.
\newblock {\em IEEE Transactions on Power Systems}, 27(2):790--799, 2012.

\bibitem{gemine2013active}
Q.~Gemine, E.~Karangelos, D.~Ernst, and B.~Corn{\'e}lusse.
\newblock Active network management: planning under uncertainty for exploiting
  load modulation.
\newblock In {\em Proceedings of the 2013 IREP Symposium - Bulk Power System
  Dynamics and Control - IX}, page~9, 2013.

\bibitem{gill2014dynamic}
S.~Gill, I.~Kockar, and G.W. Ault.
\newblock Dynamic optimal power flow for active distribution networks.
\newblock {\em Power Systems, IEEE Transactions on}, 29(1):121--131, 2014.

\bibitem{macedo2015optimal}
Leonardo~H Macedo, John~F Franco, Marcos~J Rider, and Ruben Romero.
\newblock Optimal operation of distribution networks considering energy storage
  devices.
\newblock {\em Smart Grid, IEEE Transactions on}, 6(6):2825--2836, 2015.

\bibitem{olivier2015active}
Fr{\'e}d{\'e}ric Olivier, Petros Aristidou, Damien Ernst, and Thierry
  Van~Cutsem.
\newblock Active management of low-voltage networks for mitigating overvoltages
  due to photovoltaic units.
\newblock {\em Smart Grid, IEEE Transactions on}, 7(2):926--936, 2015.

\bibitem{dommel1968optimal}
H.W. Dommel and W.F. Tinney.
\newblock Optimal power flow solutions.
\newblock {\em IEEE transactions on Power Apparatus and Systems},
  PAS-87(10):1866--1876, 1968.

\bibitem{capitanescu2007ipm}
F.~Capitanescu, M.~Glavic, D.~Ernst, and L.~Wehenkel.
\newblock Interior-point based algorithms for the solution of optimal power
  flow problems.
\newblock {\em Electric Power Systems Research}, 77(5--6):508--517, 2007.

\bibitem{jabr2006radial}
Rabih~A Jabr.
\newblock Radial distribution load flow using conic programming.
\newblock {\em Power Systems, IEEE Transactions on}, 21(3):1458--1459, 2006.

\bibitem{lavaei2012}
J.~Lavaei and S.H. Low.
\newblock Zero duality gap in optimal power flow problem.
\newblock {\em IEEE Transactions on Power Systems}, 27(1):92--107, 2012.

\bibitem{bose2012quadratically}
S.~{Bose}, D.F. {Gayme}, K.~{Mani Chandy}, and S.H. {Low}.
\newblock Quadratically constrained quadratic programs on acyclic graphs with
  application to power flow.
\newblock {\em ArXiv e-prints}, 2012.

\bibitem{gan2012branch}
L.~Gan, N.~Li, U.~Topcu, and S.~Low.
\newblock On the exactness of convex relaxation for optimal power flow in tree
  networks.
\newblock In {\em Proceedings of the IEEE 51st Annual Conference on Decision
  and Control (CDC)}, pages 465--471, 2012.

\bibitem{phan2012}
DT~Phan.
\newblock Lagrangian duality and branch-and-bound algorithms for optimal power
  flow.
\newblock {\em Operations Research}, 60(2):275--285, 2012.

\bibitem{gopalakrishnan2012global}
A.~Gopalakrishnan, A.U. Raghunathan, D.~Nikovski, and L.T. Biegler.
\newblock Global optimization of optimal power flow using a branch \& bound
  algorithm.
\newblock In {\em Proceedings of the 50th Annual Allerton Conference on
  Communication, Control, and Computing}, pages 609--616, 2012.

\bibitem{bolognani2016existence}
Saverio Bolognani and Sandro Zampieri.
\newblock On the existence and linear approximation of the power flow solution
  in power distribution networks.
\newblock {\em Power Systems, IEEE Transactions on}, 31(1):163--172, 2016.

\bibitem{gayme2011optimal}
Dennice Gayme and Ufuk Topcu.
\newblock Optimal power flow with distributed energy storage dynamics.
\newblock In {\em Proceedings of the 2011 American Control Conference (ACC)},
  pages 1536--1542, 2011.

\bibitem{gopalakrishnan2013global}
A.~Gopalakrishnan, A.U. Raghunathan, D.~Nikovski, and L.T. Biegler.
\newblock Global optimization of multi-period optimal power flow.
\newblock In {\em Proceedings of the 2013 American Control Conference (ACC)},
  pages 1157--1164, 2013.

\bibitem{alguacil2000multiperiod}
N.~Alguacil and A.J. Conejo.
\newblock Multiperiod optimal power flow using {B}enders decomposition.
\newblock {\em IEEE Transactions on Power Systems}, 15(1):196--201, 2000.

\bibitem{gemine2014relaxations}
Q.~Gemine, D.~Ernst, Q.~Louveaux, and B.~Corn{\'e}lusse.
\newblock Relaxations for multi-period optimal power flow problems with
  discrete decision variables.
\newblock In {\em Proceedings of the 18th Power Systems Computation Conference
  (PSCC-14)}, page~7, 2014.

\bibitem{andersson2004modelling}
G.~Andersson.
\newblock Modelling and analysis of electric power systems.
\newblock {\em EEH-Power Systems Laboratory, Swiss Federal Institute of
  Technology (ETH), Z{\"u}rich, Switzerland}, 2004.

\bibitem{monticelli1999state}
A.~Monticelli.
\newblock {\em State estimation in electric power systems: a generalized
  approach}, volume 507.
\newblock Springer Science \& Business Media, 1999.

\bibitem{engelhardt2011reactive}
Stephan Engelhardt, Istvan Erlich, Christian Feltes, J{\"o}rg Kretschmann, and
  Fekadu Shewarega.
\newblock Reactive power capability of wind turbines based on doubly fed
  induction generators.
\newblock {\em Energy Conversion, IEEE Transactions on}, 26(1):364--372, 2011.

\bibitem{soleimani2016receding}
Hamid Soleimani~Bidgoli, Mevludin Glavic, and Thierry Van~Cutsem.
\newblock Receding-horizon control of distributed generation to correct voltage
  or thermal violations and track desired schedules.
\newblock In {\em Proceedings of the 19th Power Systems Computation Conference
  (PSCC 2016)}, 2016.

\bibitem{powell2016unified}
W.B. Powell.
\newblock {A} {U}nified {F}ramework for {O}ptimization under {U}ncertainty.
\newblock Informs {T}ut{OR}ials in {O}perations {R}esearch, 2016.

\bibitem{bellman1957dp}
R.~Bellman.
\newblock {\em Dynamic Programming}.
\newblock Princeton University Press, 1957.

\bibitem{bertsekas1978stochastic}
D.P. Bertsekas and S.E. Shreve.
\newblock {\em Stochastic {O}ptimal {C}ontrol: {T}he {D}iscrete {T}ime {C}ase}.
\newblock Academic Press New York, 1978.

\bibitem{Shapiro2009}
A.~Shapiro, D.~Dentcheva, and A.~Ruszczy\'nski.
\newblock {\em Lectures on {S}tochastic {P}rogramming: {M}odeling and
  {T}heory}.
\newblock SIAM, 2009.

\bibitem{defourny2011multistage}
B.~Defourny, D.~Ernst, and L.~Wehenkel.
\newblock {\em Multistage stochastic programming: A scenario tree based
  approach to planning under uncertainty}, chapter~6, page~51.
\newblock Information Science Publishing, Hershey, PA, 2011.

\bibitem{fortuny1981representation}
Jos{\'e} Fortuny-Amat and Bruce McCarl.
\newblock A representation and economic interpretation of a two-level
  programming problem.
\newblock {\em Journal of the operational Research Society}, pages 783--792,
  1981.

\bibitem{ieee:2015_1}
F.~Capitanescu, L.~F. Ochoa, H.~Margossian, and N.~D. Hatzargyriou.
\newblock Assessing the potential of network reconfiguration to improve
  distributed generation hosting capacity in active distribution systems.
\newblock {\em IEEE Trans. Power Syst.}, 1:346--356, 2015.

\bibitem{ukgds}
{SEDG Centre}.
\newblock {UK} generic distribution system ({UKGDS}) project.
\newblock \url{http://www.sedg.ac.uk/}, 2010.

\bibitem{redner1984mixture}
R.A. Redner and H.F. Walker.
\newblock Mixture densities, maximum likelihood and the {EM} algorithm.
\newblock {\em SIAM Review}, 26(2):pp. 195--239, 1984.

\bibitem{bishop2006pattern}
C.M. Bishop.
\newblock {\em Pattern recognition and machine learning}.
\newblock springer, 2006.

\bibitem{marin2012approximate}
J.-M. Marin, P.~Pudlo, C.P. Robert, and R.J. Ryder.
\newblock Approximate bayesian computational methods.
\newblock {\em Statistics and Computing}, 22(6):1167--1180, 2012.

\bibitem{grelaud2009abc}
A.~Grelaud, C.P. Robert, J.-M. Marin, F.~Rodolphe, and J.-F. Taly.
\newblock {ABC} likelihood-free methods for model choice in gibbs random
  fields.
\newblock {\em Bayesian Analysis}, 4(2):317--335, 06 2009.

\bibitem{gemine16gaussian}
Quentin Gemine, Bertrand Corn{\'e}lusse, Mevludin Glavic, Rapha{\"e}l
  Fonteneau, and Damien Ernst.
\newblock A gaussian mixture approach to model stochastic processes in power
  systems.
\newblock In {\em Proceedings of the 19th Power Systems Computation Conference
  (PSCC'16)}, 2016.

\bibitem{jones2014scipy}
E.~Jones, T.~Oliphant, and P.~Peterson.
\newblock {SciPy}: Open source scientific tools for {Python}.
\newblock 2014.

\bibitem{pedregosa2011scikit}
F.~Pedregosa, G.~Varoquaux, A.~Gramfort, V.~Michel, B.~Thirion, O.~Grisel,
  M.~Blondel, P.~Prettenhofer, R.~Weiss, V.~Dubourg, and al.
\newblock Scikit-learn: Machine learning in python.
\newblock {\em The Journal of Machine Learning Research}, 12:2825--2830, 2011.

\bibitem{hastie2005elements}
Trevor Hastie, Robert Tibshirani, Jerome Friedman, and James Franklin.
\newblock The elements of statistical learning: data mining, inference and
  prediction.
\newblock {\em The Mathematical Intelligencer}, 27(2):83--85, 2005.

\bibitem{hart2012pyomo}
W.E. Hart, C.~Laird, J.-P. Watson, and D.L. Woodruff.
\newblock {\em Pyomo--optimization modeling in python}, volume~67.
\newblock Springer Science \& Business Media, 2012.

\bibitem{bonmin}
P.~Bonami and J.~Lee.
\newblock {BONMIN} users manual.
\newblock Technical report, 2006.

\bibitem{busoniu2011cross}
L.~Busoniu, D.~Ernst, B.~De~Schutter, and R.~Babuska.
\newblock Cross-entropy optimization of control policies with adaptive basis
  functions.
\newblock {\em IEEE Transactions on Systems, Man, and Cybernetics, Part B:
  Cybernetics}, 41(1):196--209, 2011.

\bibitem{bertsekas1996neuro}
D.P. Bertsekas and J.N. Tsitsiklis.
\newblock {\em {N}euro-{D}ynamic {P}rogramming}.
\newblock Athena Scientific, Belmont, MA, 1996.

\bibitem{busoniu2010reinforcement}
L.~Busoniu, R.~Babuska, B.~De~Schutter, and D.~Ernst.
\newblock {\em Reinforcement {L}earning and {D}ynamic {P}rogramming {U}sing
  {F}unction {A}pproximators}.
\newblock CRC Press, Boca Raton, FL, 2010.

\bibitem{powell2014clearing}
W.B. Powell.
\newblock {C}learing the {J}ungle of {S}tochastic {O}ptimization.
\newblock Informs {T}ut{OR}ials, 2014.

\end{thebibliography}

\end{document}